\documentclass[superscriptaddress,twocolumn,secnumarabic,amssymb,amsmath,nobibnotes,aps,prd,showpacs,nofootinbib]{revtex4}
\usepackage{graphicx}
\usepackage{epsf}
\usepackage{bm}
\usepackage{amsmath}
\usepackage{amsfonts}
\usepackage{amssymb}
\usepackage{epstopdf}%
\providecommand{\U}[1]{\protect\rule{.1in}{.1in}}

\newcommand{\be}{\begin{equation}}
\newcommand{\ee}{\end{equation}}

\newcommand{\mincir}{\raise
-3.truept\hbox{\rlap{\hbox{$\sim$}}\raise4.truept\hbox{$<$}\ }}
\newcommand{\magcir}{\raise
-3.truept\hbox{\rlap{\hbox{$\sim$}}\raise4.truept\hbox{$>$}\ }}

\usepackage{color}

\begin{document}

\title{Forecasting Interacting Vacuum-Energy Models using Gravitational Waves}

\author{Weiqiang Yang}
\email{d11102004@163.com}
\affiliation{Department of Physics, Liaoning Normal University, Dalian, 116029, P. R. China}

\author{Supriya Pan}
\email{supriya.maths@presiuniv.ac.in}
\affiliation{Department of Mathematics, Presidency University, 86/1 College Street, Kolkata 700073, India}

\author{Eleonora Di Valentino}
\email{eleonora.divalentino@manchester.ac.uk}
\affiliation{Jodrell Bank Center for Astrophysics, School of Physics and Astronomy, University of Manchester, Oxford Road, Manchester, M13 9PL, UK}

\author{Bin Wang}
\email{wang_b@sjtu.edu.cn}
\affiliation{School of Aeronautics and Astronautics, Shanghai Jiao Tong University, Shanghai 200240, P. R. China}
\affiliation{Centre for Gravitation and Cosmology, Yangzhou University, Yangzhou 225009, P. R. China}

\author{Anzhong Wang}
\email{Anzhong_Wang@baylor.edu}
\affiliation{Institute for Theoretical Physics $\&$ Cosmology,
Zhejiang University of Technology, Hangzhou, 310023, P. R. China}
\affiliation{GCAP-CASPER, Department of Physics, Baylor University, Waco, TX, 76798-7316, USA}

\begin{abstract}
The physics of the dark sector has remained one of the controversial areas of modern cosmology at present and hence it naturally attracts massive attention to the scientific community. With the developments of the astronomical data, the physics of the dark sector is becoming much more transparent than it was some twenty years back. The detection of gravitational waves (GWs) 
has now opened a cluster of possibilities in the cosmological regime. Being motivated by the detection of GWs and its possible impact on the physics of dark matter and dark energy, in this work we focus on the interacting dark energy models. Assuming the simplest possibility in which the vacuum energy with equation-of-state $w_x =-1$ is allowed to interact with the pressureless dark matter, we have extracted the constraints of the cosmological parameters.  We find that the addition of the GWs data to the CMB measurements significantly improves up to a factor 4 of the parameters space, and up to a factor 2 for the full combination of current cosmological datasets, namely CMB+BAO+Pantheon+RSD+R16+CC+WL. The most affected parameters by the inclusion of the GWs are 
$\Omega_ch^2$, $\theta_{MC}$, $\xi$, and the derived parameters $\Omega_{m0}$, $\sigma_8$ and $H_0$. 

\end{abstract}

\maketitle
\section{Introduction}

We have already crossed 20 years of the detection of the late time accelerating phase of the universe \cite{snia1, snia2}. During this period, the cosmology has witnessed a rapid change in its development, however, we are looking for a theory that could explain the present observational data in a satisfactory way. The candidate, widely known as dark energy (DE) to the scientific community, has remained dark even after a number of promising observational tests. According to the present observational evidences, nearly 68\% of the total energy density of the universe is filled up with such DE and around 28\% of the total energy density of the universe is  filled up with another dark fluid, namely dark matter (DM) \cite{Ade:2015xua,Aghanim:2018eyx}. The dynamics of them is also not yet understood to us. The modelling of the universe has thus been greatly dependent on the evolution of these dark components. Based on the current literature, a number of cosmological models are already existing aiming to provide a clear description of the universe's evolution in agreement with the observational data from various sources. 
These cosmological models can be classified into two broad classes, one is known as the non-interacting cosmological models in which both the dark fluids, namely DM and DE, enjoy independent evolution. On the other hand, the remaining class of models is known as the interacting cosmological models in which DM and DE interact with each other through matter flow between them.

The interacting cosmological models have gained a massive attention in the cosmological community for explaining many observational issues related to the dynamics of the universe \cite{He:2008tn,He:2009mz,He:2010ta,Yang:2014gza,Yang:2014hea,Pan:2012ki,Li:2015vla,Yang:2016evp,Pan:2016ngu,Sharov:2017iue,Shahalam:2017fqt,Guo:2017hea,Cai:2017yww,Yang:2017yme,Pan:2017ent,Yang:2018xlt,Yang:2018ubt,Yang:2018pej,vonMarttens:2018iav,Yang:2018qec,Asghari:2019qld,Zhang,Paliathanasis:2019hbi,Pan:2019jqh} (see \cite{Bolotin:2013jpa, Wang:2016lxa} for  extensive reviews on interacting dark energy theory). 
The theory of interaction has been popular in order to explain the cosmic coincidence problem \cite{Amendola-ide1,Amendola-ide2,Pavon:2005yx,delCampo:2008sr,delCampo:2008jx}. According to the recent 
observational evidences, the interaction between DM and DE, although very mild (this mostly depends on the specific coupling), however,  within $1\sigma$ confidence region, we cannot rule it out \cite{Costa:2013sva,Salvatelli:2014zta,Nunes:2016dlj,Yang:2016evp,Abdalla:2014cla,Yang:2017ccc,Yang:2017zjs}.  Additionally, the investigations in this direction directly show that even if a mild interaction in the dark sector could solve the tensions in various cosmological parameters arising from their local and global predictions, including the tension on the Hubble constant $H_0$ \cite{Kumar:2017dnp,DiValentino:2017iww,Yang:2018euj,Yang:2018uae,Kumar:2019wfs} and the tension on the amplitude of the matter power spectrum $\sigma_8$ \cite{Kumar:2019wfs,vandeBruck:2017idm,Thomas,An:2017crg}. Additionally, the dark sector's interaction can also explain an excess amount of 21 cm absorption signal around the redshift $z \sim 17$ that has been recently detected by the Experiment to Detect the Global Epoch of  Reionization  Signature (EDGES) \cite{Costa:2018aoy,Xiao:2018jyl}. 
Apart from that, theory of interaction further includes some other interesting aspects in which the crossing of the phantom divide line is one of them \cite{Wang:2005jx,Pan:2014afa}. Therefore, based on the above observations, both from theoretical and observational grounds, interacting dark energy theory might be considered to be a potential area for further investigations.  

The current work has thus been motivated to work in this area while the main  incitement of this work is not only to investigate the interacting models with the standard cosmological data, but also we proceed one step further by investigating this area using the simulated gravitational waves (GWs) data from the Einstein Telescope. The use of GWs has potential implications in cosmological studies that we shall point out here. As already known, GWs have been detected by LIGO and VIRGO collaborations in recent times \cite{ligo01,Abbott:2016nmj,Gw03,Gw04,Gw05, Gw06, Gw07} with a massive thrilling  in the entire scientific community. Probably this is one of the greatest achievements of modern science with a potential indication for a new era in cosmology and astrophysics. In fact GWs have another important aspect that comes through the understanding of primordial gravitational waves, that means the detection of gravitational waves in the extreme early phase of the universe. From the effects of  primordial GWs on the cosmic microwave background radiation, the origin of the universe can better be understood. Naturally, through the analysis of GWs, it is expected to have new cosmological and astrophysical information. Following the detection of GWs the screening of cosmological models has already been started by many investigators that includes analysis of various cosmological models \cite{Creminelli:2017sry,Ezquiaga:2017ekz,Sakstein:2017xjx, Chakraborty:2017qve,Visinelli:2017bny,Oost:2018tcv,Casalino:2018tcd,Nunes:2018evm,Nunes:2018zot,Mifsud:2019hjw,Qi:2019wwb,Palmese:2019ehe} as well as the estimation of the Hubble constant \cite{DiValentino:2017clw,DiValentino:2018jbh}. In summary from the investigations and analysis carried out in the recent time using the GWs data, it is quite reasonable to test the effects of GWs on other areas of cosmology. Since the interaction in the dark sector has remained one of the most talkative issues since quite a long time in the cosmological history, and it is one of the promising research areas  to understand the tensions in the cosmological parameters, thus, undoubtedly one should apply GWs to this particular zone aiming to look for more precise conclusion in this direction. Now, following this motivation in the present work we have considered an interacting scenario where cold dark matter  directly interacts with vacuum. We consider two well motivated interaction models and analyze the scenarios using (i) the standard cosmological data such as, cosmic microwave background radiation (CMB), baryon acoustic oscillations (BAO), recent Pantheon sample of supernovae type Ia, redshift space distortion (RSD), local measurement of Hubble constant (R16), Hubble parameter measurements from cosmic chronometers (CC), weak gravitational lensing (WL),  and then (ii) we include the simulated GWs data from Einstein Telescope with those standard cosmological datasets. 
By comparing the results from two distinct analyses, it shows that the inclusion of GWs data significantly improves the parameter space of the interactive models compared to their constraints coming from the usual cosmological datasets. This is an interesting outcome of the present work and possibly will motivate other investigators to continue the same using the GWs.

The work is organized in the following manner. In section \ref{sec2} we present the gravitational field equations in the presence of an interaction in the dark fluids. The section \ref{sec-data} divided into two subsections where in the first subsection \ref{subsec-data} we describe the current standard cosmological data, while in the subsection \ref{GWdata} we describe the method to simulate the gravitational waves data. After that in section \ref{sec-results} we present the results of the analysis of the interacting models.  Finally, in section \ref{sec-summary} we end the present work with a brief summary of all the findings.

\section{The interacting universe: Gravitational equations}
\label{sec2}

We consider an interacting scenario between DM and DE where the underlying gravitational sector follows the Einstein gravity. The conservation equation for the combined dark sector, that is, DM plus DE, follows, 

\begin{eqnarray}
\nabla^\mu (T^{DM}_{\mu\nu} + T^{DE}_{\mu\nu})= 0,\label{conservation}
\end{eqnarray}
which for  a Friedmann-Lema\^{i}tre-Robertson-Walker line element,

\begin{eqnarray}
ds^2 = -dt^2 + a^2 (t) \left[\frac{dr^2}{1-kr^2} + r^2 \left(d \theta^2 + \sin^2 \theta d \phi^2 \right) \right],
\end{eqnarray}
where $a(t)$ denotes  the expansion scale factor of the universe, and $k$ is the spatial curvature of the universe\footnote{We note that $k$ could take three different values, namely, $k =0$ (flat universe), $k =-1$ (open universe) and $k =+1$ (closed universe). }, can be decoupled into 

\begin{eqnarray}
\dot{\rho}_x  = Q, \label{cons1}\\
\dot{\rho}_c + 3 H \rho_c = - Q,\label{cons2}
\end{eqnarray}
where an overhead dot represents the cosmic time derivative; 
$H \equiv \dot{a}/a$ is the Hubble rate of the FLRW universe; $\rho_c$, $\rho_x$ are respectively the energy density of the  pressureless dark matter and the vacuum.  The quantity $Q$ refers to the interaction rate between vacuum and the cold dark matter. In general there is no such specific rule to select any particular interaction rate, thus, usually we allow some phenomenological models for $Q$ and test the underlying cosmological scenario with the observations. This essentially reconstructs  the expansion history of the universe in the presence of the interaction. In addition to that we assume the presence of baryons and radiation in the universe  but both of them have their own conservation equation. That means, under their usual equation of state, $w_i = p_i/\rho_i$ ($i = r, b$ where $r$ is for radiation and $b$ for baryons), their evolution equations yield, $\rho_r = \rho_{r0} a^{-4}$, $\rho_b  = \rho_{b0} a^{-3}$. Here $\rho_{r0}$, $\rho_{b0}$, respectively denote the present energy density of radiation and baryons. The only constraint in such a universe is 
\begin{eqnarray}\label{Hubble}
H^2 + \frac{k}{a^2} = \frac{8 \pi G}{3} \sum \rho_i,
\end{eqnarray}
however, throughout the present work we shall consider the spatially flat case, that means we assume $k =0$. Now, we comment that, for any given interaction function $Q$, the conservation equations (\ref{cons1}) and (\ref{cons2}) together with the Hubble function (\ref{Hubble}) can determine the dynamics of the interacting universe.  

Let us now move to the cosmological perturbations of the interacting vacuum models. In what follows we consider the following perturbed FLRW metric \cite{Mukhanov, Ma:1995ey, Malik:2008im}

\begin{eqnarray}\label{perturbed-metric}
ds^{2}=a^{2}(\tau )\Bigg[-(1+2\phi )d\tau ^{2}+2\partial _{i}Bd\tau dx^{i}\notag\\+
\Bigl((1-2\psi )\delta _{ij}+2\partial _{i}\partial _{j}E\Bigr)dx^{i}dx^{j}%
\Bigg],
\end{eqnarray}
where $\tau$ is the conformal time and  
$\phi $, $B$, $\psi $, $E$ are
the gauge-dependent scalar perturbation quantities. Now, using the perturbed metric (\ref{perturbed-metric}), and the conservation equations
\begin{equation*}\label{conservation-1}
\nabla _{\nu }T_{A}^{\mu \nu }=Q_{A}^{\mu },~~\sum\limits_{\mathrm{A}}{Q_{A}^{\mu }}=0,
\end{equation*}%
one can calculate the gravitational field equations, see \cite{Majerotto:2009np, Valiviita:2008iv, Clemson:2011an}. We note that the above equation $A \in \{c, x \}$ where $A = c$, is for CDM and $A =x$ for vacuum energy.  

Here, we shall work in the synchronous gauge, that means, $\phi =B=0$, $\psi =\eta $, and $k^{2}E=-h/2-3\eta $, with $k$ being the Fourier mode and $h$, $\eta$ are the metric perturbations. We also introduce some notations that will be useful later. We introduce $\delta _{A}=\delta \rho _{A}/\rho_{A}$, as the density perturbations for the fluid $A$, and $\theta  =  \theta_{\mu}^{\mu}$ is the volume expansion scalar of the total fluid, thus, for $\theta_c$ we mean the volume expansion scalar for the CDM fluid. Thus, with all the above notations in hand, in the synchronous gauge, the momentum conservation equation for CDM is reduced to  \cite{Wang:2014xca}: $\dot{\theta}_c = 0.$

The density perturbations for CDM in the comoving synchronous gauge can be found as \cite{Wang:2014xca} 
\begin{eqnarray}
\dot{\delta}_c = -\frac{\dot{h}}{2} + \frac{Q}{\rho_c} \delta_c~.
\end{eqnarray}
We note that in this gauge, the vacuum energy is spatially homogeneous which means that, $\delta \rho_x = 0$.  Thus, having both the background and perturbative evolutions for any coupling function $Q$, we are now able to proceed with the examinations of the models.

Hence, with all of  these above, we now  close this section with the models of interaction that we study here. In what follows we propose the following two models for $Q$: 

\begin{eqnarray}
&&{\rm Model\; 1:}\; Q = 3 H \xi \rho_{x} (t),\label{ivs1}\\
&&{\rm Model\; 2:}\;Q = 3 H \xi \rho_c (t) \rho_{x} (t) (\rho_c (t) +\rho_{x} (t))^{-1},\label{ivs2}
\end{eqnarray}
where in both the expressions of $Q$, $\xi$
refers to the coupling parameter (sometimes it is called the coupling strength) of the models. The coupling parameter, $\xi$, has the following two properties: (i) the strength of the interaction via the magnitude of $\xi$ and (ii) the direction of energy flow between the dark sectors through its sign. Let us note that the cosmological scenario with the interaction function (\ref{ivs1}) (i.e. {\rm Model\; 1}) is labeled as IVS1 (Interacting Vacuum Scenario 1) and the cosmic scenario with {\rm Model\; 2} (i.e. eqn. (\ref{ivs2})) is labeled as IVS2 (Interacting Vacuum Scenario 2).

\begin{table}
\begin{center}
\begin{tabular}{c|c}
Parameter                    & Prior\\
\hline 
$\Omega_{b} h^2$             & $[0.005,0.1]$\\
$\Omega_{c} h^2$             & $[0.01,0.99]$\\
$\tau$                       & $[0.01,0.8]$\\
$n_s$                        & $[0.5, 1.5]$\\
$\log[10^{10}A_{s}]$         & $[2.4,4]$\\
$100\theta_{MC}$             & $[0.5,10]$\\ 
$\xi$                        & $[-1, 1]$\\
\end{tabular}
\end{center}
\caption{This table describes the flat priors on the model parameters that we have used during the statistical analysis. Here, $\Omega_bh^2$, $\Omega_{c}h^2$, are the baryon and cold dark matter densities, respectively;  $\tau$ refers to the reionization optical depth; $100 \theta_{MC}$ denotes the ratio of sound horizon to the angular diameter distance; $n_s$ is the scalar spectral index; $A_S$ is the amplitude of the primordial scalar power spectrum and finally $\xi$ denotes the coupling parameter of the interaction models (\ref{ivs1}) and (\ref{ivs2}). }
\label{tab:priors}
\end{table}

\section{Observational data}
\label{sec-data}

In this section we describe the observational data used in the analysis dividing them into two subsections: one for the current cosmological probes and the other for the simulated gravitational waves data from the Einstein Telescope. 

\subsection{Current cosmological probes}
\label{subsec-data}

\begin{itemize}

\item CMB: The Cosmic Microwave Background (CMB) data are one of the potential astronomical probes to analyze the DE models. Here, we make use of the full range of multipoles in temperature and polarization of the CMB angular power spectra from the Planck satellite (identified as Planck TT, TE, EE + lowTEB)  \cite{Adam:2015rua, Aghanim:2015xee}. 

\item SNIa: The supernovae Type Ia (SNIa) data were the first observational data reporting the accelerating phase of the universe. With the developments of the observational data, different versions with significant compilation of the SNIa had been available. In the present work we make use of the latest  Pantheon sample \cite{Scolnic:2017caz}. 

\item BAO: We include the Baryon Acoustic Oscillations (BAO) data from different observational missions \cite{Beutler:2011hx, Ross:2014qpa,Gil-Marin:2015nqa}.

\item RSD: The Redshift Space Distortion (RSD) data are included in the analysis \cite{Gil-Marin:2016wya}.

\item Hubble: We also include the Hubble parameter measurements from the Cosmic Chronometers (CC) \cite{Moresco:2016mzx}.

\item R16: We include the local measurement of the Hubble constant value yielding $H_0=73.24\pm1.74$ km/s/Mpc at $68 \%$ CL \cite{Riess:2016jrr}. 

\item WL: Data from weak lensing (WL) are also important for dark energy analysis. Here we use the cosmic shear data from the blue galaxy sample compiled from the Canada-France-Hawaii Telescope Lensing Survey (CFHTLenS) \cite{Heymans:2013fya,Asgari:2016xuw}.

\end{itemize}

\subsection{Gravitational waves data: Method of simulation}
\label{GWdata}

Here we describe the mechanism to 
simulate the Gravitational Waves Standard Sirens (GWSS) data in which each data point consists of the following triplet $(z, d_{{L}} (z), \sigma_{d_{L}})$ of a GW source: $z$ is the redshift at which the measurement is performed; $d_{L} (z)$ denotes  the luminosity distance at redshift $z$; $\sigma_{d_{L}}$ is the corresponding error.

The generation of the GWSS data is the initial step of this work. This is performed by simulating the redshift distribution of the sources with the assumption that the redshifts of all observed GW sources are available to us. We focus on GW events that originate from 2 distinct types of binary systems: (I) the binary system combining a Black Hole (BH) and a Neutron Star (NS) identified as BHNS and (II) the binary neutron star (BNS). Let us come to the main mathematical part of the analysis.

In order to proceed we need that redshift distribution of the observable sources which follows \cite{Zhao:2010sz,Cai:2016sby,Wang:2018lun}

\begin{equation}
P(z)\propto \frac{4\pi d_C^2(z)R(z)}{H(z)(1+z)},
\label{equa:pz}
\end{equation}
where $d_C(z)$ is  the co-moving distance at redshift $z$ and $R(z)$ is the merger rate of binary systems that may include either BHNS or BNS with 
\cite{Cai:2016sby,Schneider:2000sg,Cutler:2009qv}

\begin{equation}
R(z)=\begin{cases}
1+2z, & z\leq 1, \\
\frac{3}{4}(5-z), & 1<z<5, \\
0, & z\geq 5.
\end{cases}
\label{equa:rz}
\end{equation}

Now, we come to the technical part of the GW simulation.  The complete configuration of our simulation is based on the prediction of the Advanced LIGO-Virgo network.  
The ratio between the observed binary events (BHNS and BNS) is fixed to be $0.03$ that makes BNS the profouse majority of GW sources. Considering a rough approximation of the mass distribution of the astrophysical objects, namley, neutron stars and black holes, we perform a random sampling of their masses from uniform distributions $U(M_{\odot}, 2 M_{\odot})$ and $U(3 M_{\odot}, 10 M_{\odot})$ respectively. Let us note that $M_{\odot}$ represents one solar mass. For a detailed information and discussions in this direction one can see the references \cite{Cai:2016sby,Wang:2018lun}.

Hence, following the approach described above one can obtain the catalogue of the GWSS data by introducing the fiducial model. The fiducial model could be any well motivated cosmological model in principle. Now, for any fiducial model, in the background of a FLRW universe, one can be able to find the 
luminosity distance $d_L (z)$ using 

\begin{equation}
d_L (z) = {(1 + z)}\int_0^z {\frac{{dz'}}{{H(z')}}}.
\label{equa:dl}
\end{equation}
where $H (z)$ refers to the Hubble function corresponding to the fiducial model.  Therefore,  using Eq.~(\ref{equa:dl}), one can compute the mean luminosity distances of all the GW sources which means that a relation between $d_{L} (z)$ vs. $z$ can be found for the underlying fiducial model.

Usually, $\Lambda$CDM is considered to be the fiducial model, but however, one may consider some other cosmological model as well in order to  generate the simulated GW data as technically there is no such binding to choose any other model. Here we have not fixed $\Lambda$CDM as the fiducial model which is generally considered for its simplicity (see \cite{Zhao:2010sz,Cai:2016sby}), but we have considered our interacting scenarios as the fiducial model. In particular, for each combination of datasets and models considered, we used the obtained best fit for simulating the GW data, and we forecast the improvement obtained on the constraints by the addition of them.

Now, once the luminosity distance of the GW source is known, it is essential to calculate the error associated with it, denoted by $\sigma_{d_{L}}$. The determination of the  error demands the expression of the GWs signal, that means, the strain of GW interferometers. Since the amplitude of the GW depends on $d_{L} (z)$, one can extract the information concerning the luminosity distance $d_{L} (z)$ provided the other parameters, such as the masses of the underlying binary system, are evaluated from the waveform. That is why the GW events are frequently referred to as the standard sirens, (similar to the Supernovae Type Ia). Consequently, the error of GW detection (given in terms of GW SNR) is passed to $\sigma_{d_{L} (z)}$ through the Fisher matrix.

We now describe the strain of GW interferometers which is supposed to be the main part of this section.
The strain $h(t)$ in the GW interferometers assuming the transverse-traceless (TT) gauge, can be written following   \cite{Cai:2016sby,Wang:2018lun} as 
\begin{equation*}
h(t)=F_+(\theta, \phi, \psi)h_+(t)+F_\times(\theta, \phi, \psi)h_\times(t),
\end{equation*}
in which the involved quantities have the following meanings: $F_{+}$ and $F_{\times}$ are the beam pattern functions of the Einstein Telescope (ET); $\psi$ denotes the polarization angle; $\theta$, $\phi$ describe the location of the GW source relative to the GW detector (here Einstein Telescope); $h_{+} = h_{xx} = -h_{-yy}$, $h_{\times} = h_{xy} = h_{yx}$ (two independent components of the GWs tensor $h_{\mu \nu}$ in the transverse-traceless (TT) gauge). We refer to \cite{Cai:2016sby} for a detailed description about this. 

The antenna pattern functions of the Einstein Telescope can be written as \cite{Zhao:2010sz,Cai:2016sby,Wang:2018lun}

\begin{align}
F_+^{(1)}(\theta, \phi, \psi)=&~~\frac{{\sqrt 3 }}{2}\Bigl[\frac{1}{2}(1 + {\cos ^2}(\theta ))\cos (2\phi )\cos (2\psi ) \nonumber \\
&~~- \cos (\theta )\sin (2\phi )\sin (2\psi )\Bigr],\label{apf1}\\
F_\times^{(1)}(\theta, \phi, \psi)=&~~\frac{{\sqrt 3 }}{2}\Bigl[\frac{1}{2}(1 + {\cos ^2}(\theta ))\cos (2\phi )\sin (2\psi ) \nonumber\\
&~~+ \cos (\theta )\sin (2\phi )\cos (2\psi )\Bigr].\label{apf2}
\end{align}
 
Concerning the remaining two interferometers, their antenna pattern functions can be found with the use of (\ref{apf1}), (\ref{apf2}) and by substituting $\phi$ by either $\phi+ 120^\circ$ or $\phi+240^\circ$. The reason in doing so is that the three interferometers actually form an equilateral triangular shape, and therefore they make $60^\circ$ with each other.

Now following \cite{Zhao:2010sz,Li:2013lza} we compute the Fourier transform $\mathcal{H}(f)$ of the time domain waveform $h(t)$ with the assumption of stationary phase approximation. And this leads us to have, 
$\mathcal{H}(f)=\mathcal{A}f^{-7/6}\exp[i(2\pi ft_0-\pi/4+2\psi(f/2)-\varphi_{(2.0)})],$
in which $\mathcal{A}$ represents the Fourier amplitude having 
\begin{align}
\mathcal{A}=&~~\frac{1}{d_L}\sqrt{F_+^2(1+\cos^2(\omega))^2+4F_\times^2\cos^2(\omega)}\nonumber\\
&~\times \sqrt{5\pi/96}\pi^{-7/6}\mathcal{M}_c^{5/6}. \label{FA} 
\end{align}
Here, $\mathcal{M}_c$ is named as the `chirp mass' related to the the total mass $M$ ($= m_1+m_2$) of the coalescing binary system in which $m_1$, $m_2$ are the component masses. The symmetric mass ratio (SMR) is given by $\eta=m_1 m_2/M^2$ which is related to the chirp mass as $\mathcal{M}_c =M \eta^{3/5}$.
Let us mention clearly that the masses we mention here, namely $M_c$, $M$, are the observed masses obeying a relation to the intrinsic masses given by  $M_{\rm obs}=(1+z)M_{\rm int}$. This relation exhibits an enhancement of a factor by $(1 + z)$.
Let us clarify some more points related to the above equation (\ref{FA}).  With the line of sight the angle of inclination of the binary's orbital angular momentum is denoted by $\omega$, see eqn. (\ref{FA}). 

As the short gamma ray bursts are usually strongly beamed, from the coincidence observations of the short gamma ray bursts, it is indicated that the binaries should be aligned in a definite way so that  $\omega \simeq 0$ having its maximal inclination about $\omega=20^\circ$. 

Now we come to the signal-to-noise ratio (SNR) which can be calculated once the waveform of GWs is known and it is a very important quantity because in detecting the GW event it plays a very crucial role. Actually, a GW detection is confirmed if the combined SNR is found to be of at least $8$ in the Einstein Telescope   \cite{ET,Sathyaprakash:2012jk} (see for more details ~\cite{Cai:2016sby,Zhao:2010sz,Cai:2017aea,Yang:2017bkv}). The combined SNR for the network that includes three independent interferometers is 
\begin{eqnarray}\label{combinedSNR}
\rho=\sqrt{\sum\limits_{i=1}^{3}(\rho^{(i)})^2},\,\,
\mbox{where}\,\, \rho^{(i)}=\sqrt{\left\langle \mathcal{H}^{(i)},\mathcal{H}^{(i)}\right\rangle}, 
\end{eqnarray}
where inner product used in (\ref{combinedSNR}) follows \cite{Zhao:2010sz,Cai:2016sby,Wang:2018lun}
\begin{equation}
\left\langle{a,b}\right\rangle=4\int_{f_{\rm lower}}^{f_{\rm upper}}\frac{\tilde a(f)\tilde b^\ast(f)+\tilde a^\ast(f)\tilde b(f)}{2}\frac{df}{S_h(f)},
\label{euqa:product}
\end{equation}
in which the symbol `$\sim$'
refers to Fourier transformations of the corresponding quantities and  $S_h(f)$ denotes the one-side noise power spectral density. Let us mention that $S_h(f)$ has been taken to be same as in article  \cite{Zhao:2010sz}.

Using the Fisher matrix approach, one can determine the instrumental error on $d_{L}$ through 
\begin{align}\label{error}
\sigma_{d_L}^{\rm inst}\simeq \sqrt{\left\langle\frac{\partial \mathcal H}{\partial d_L},\frac{\partial \mathcal H}{\partial d_L}\right\rangle^{-1}},
\end{align}
where $\mathcal{H}$ is the GW waveform; as $\mathcal{H} \propto 1/d_L$, so we have $\partial \mathcal{H}/\partial d_L = -h/d_L$ and plugging this quantity to (\ref{error}), the instrumental error becomes \cite{Li:2013lza}:  $\sigma_{d_L}^{\rm inst}\simeq d_L/\rho$, where as already mentioned earlier, $\rho$ is the SNR. It is now important to mention that while dealing with the GW scenarios, $d_L$ is not uncorrelated with other parameters. In fact, to come up with the expression for $\sigma_{d_L}^{\rm inst}$, shown just above, the correlations with other GW parameters have been ignored, which should not be the case 
indeed. Thus, it is mandatory to include the correlations into the current framework, and in order to do this, let us note the maximal effect of the inclination on the SNR which is a factor of $2$ (between the source being face on ($\omega = 0^{\circ}$) and edge on ($\omega = 90^{\circ}$)) \cite{Li:2013lza}. Thus, in accounting the correlation between $d_L$ and $\omega$, we double the estimation of the error on $d_L$, that means effectively, the error becomes, $\sigma_{d_L}^{\rm inst}\simeq 2 d_L/\rho$. For more discussions in this direction we refer to Ref. 
\cite{Li:2013lza}.

We consider the maximal effect of inclination on the SNR which is a factor of $2$ between $\omega = 0^{\circ}$ and $\omega = 90^{\circ}$.  Now, to furnish the 
estimation of ability of the GWs data in the context of cosmological model building and their parameter estimation, we follow \cite{Li:2013lza} where the estimation of the error imposed on the luminosity is doubled, that means, $\sigma_{d_L}^{\rm inst}\simeq \frac{2d_L}{\rho}$. Furthermore, 
under the short-wave approximation, GWs are lensed in a similar way as we find with the electromagnetic waves during propagation. This results in an additional weak lensing error as  $\sigma_{d_L}^{\rm lens}$ = $0.05z d_L$ \cite{Cai:2016sby}. As a  result of that
the combined error can be calculated to be 
$\sigma_{d_L} = \sqrt{(\sigma_{d_L}^{\rm inst})^2+(\sigma_{d_L}^{\rm lens})^2}$, where
the errors $\sigma_{d_L}^{\rm inst}$ and  $\sigma_{d_L}^{\rm lens}$, are already defined above.

Henceforth, following the above methodology, 
we can generate the simulated GWSS dataset comprising of ($z$, $d_{L} (z)$, $\sigma_{d_{L}(z)}$) in which $d_L (z)$, $\sigma_{d_{L}(z)}$ are respectively the luminosity distance and its error at a particular redshift $z$. Usually one can increase the number of GW events. However, as marked in \cite{Cai:2016sby},
the sensitivity of at least 1000 GW events is approximately similar to the Planck's constraining ability. Therefore, we have considered 1000 GW events for analyzing the present cosmological models in this work. 

\begin{figure*}
    \centering
    \includegraphics[width=0.43\textwidth]{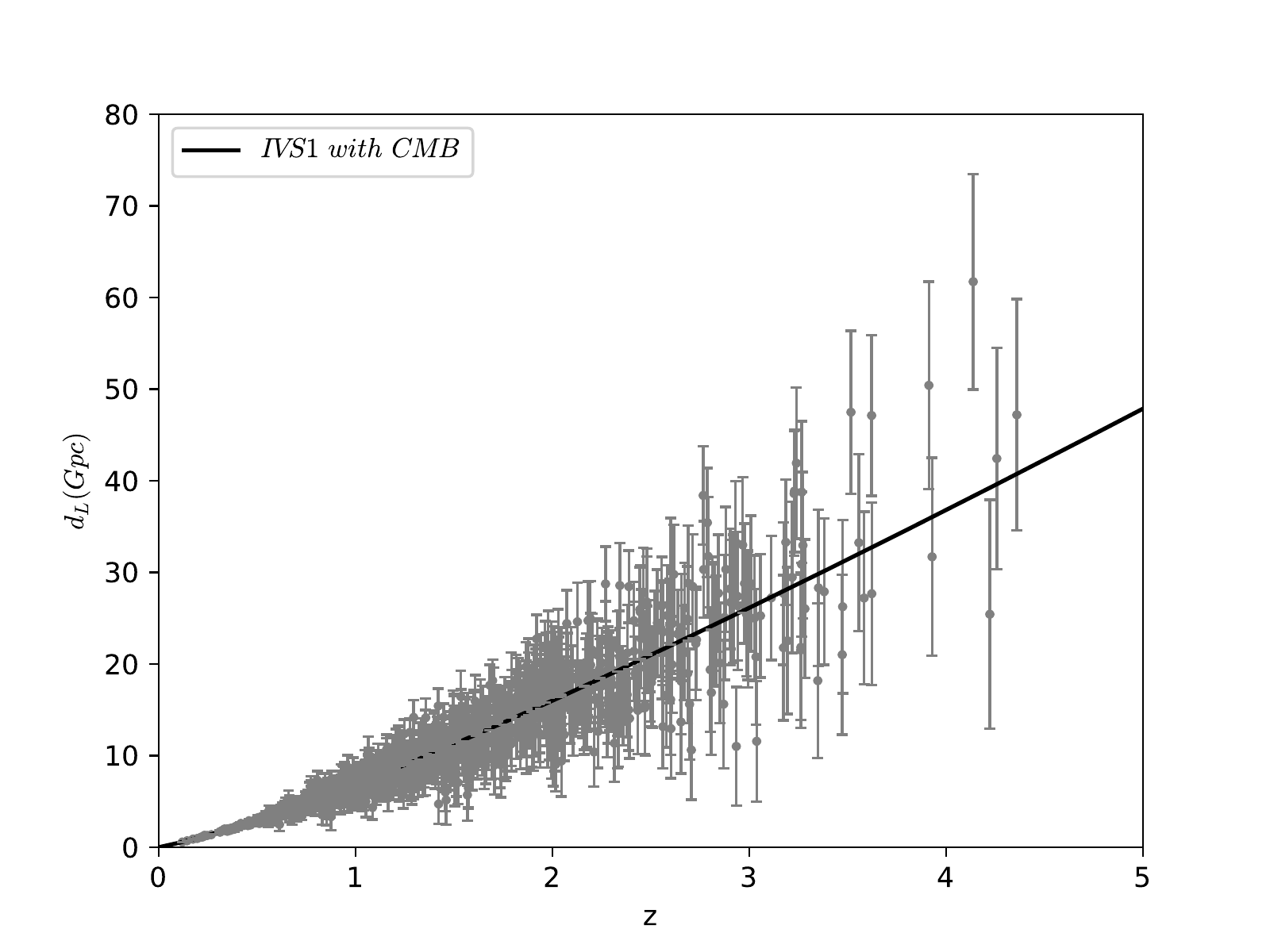}
    \includegraphics[width=0.43\textwidth]{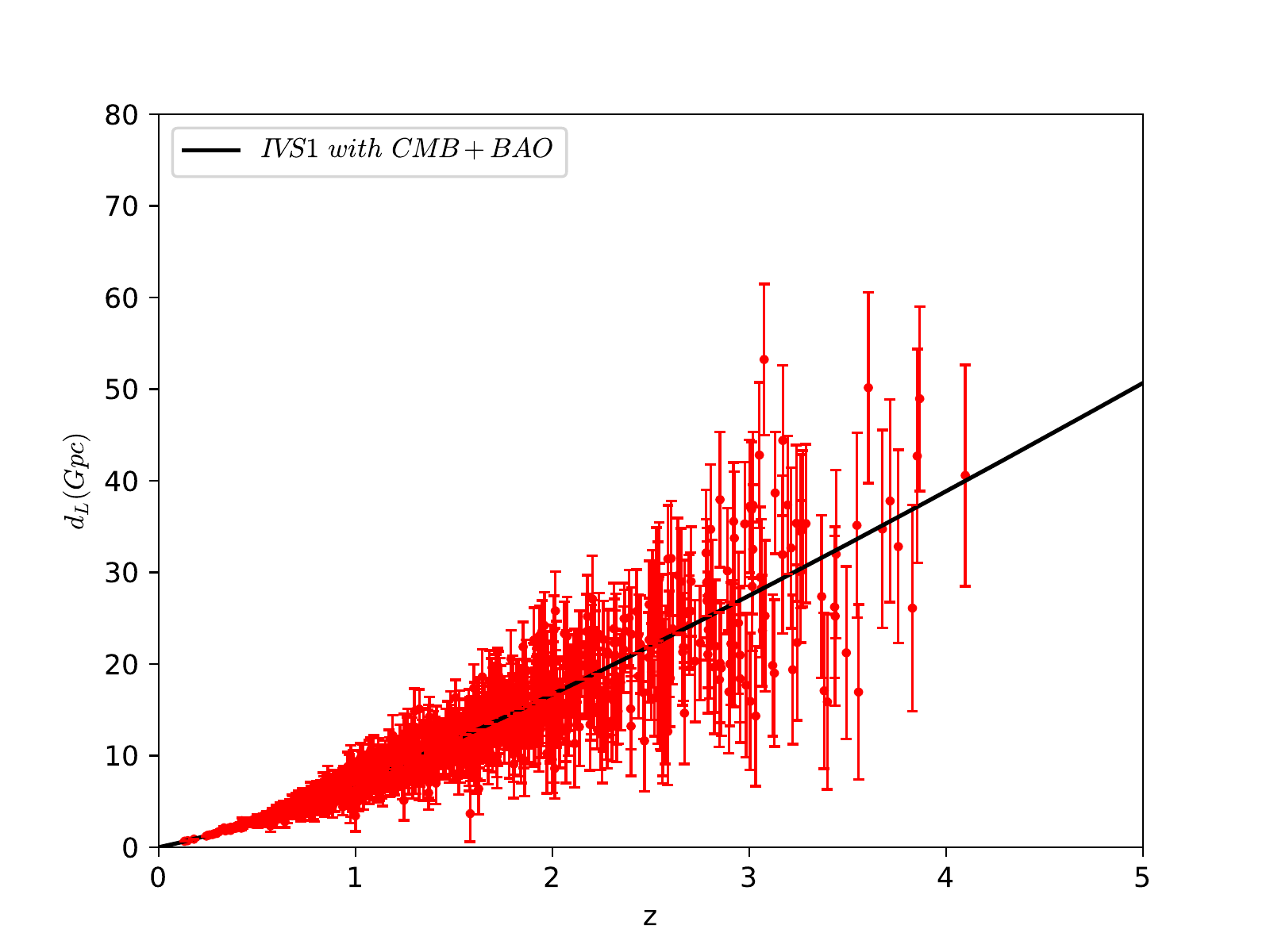}
    \includegraphics[width=0.43\textwidth]{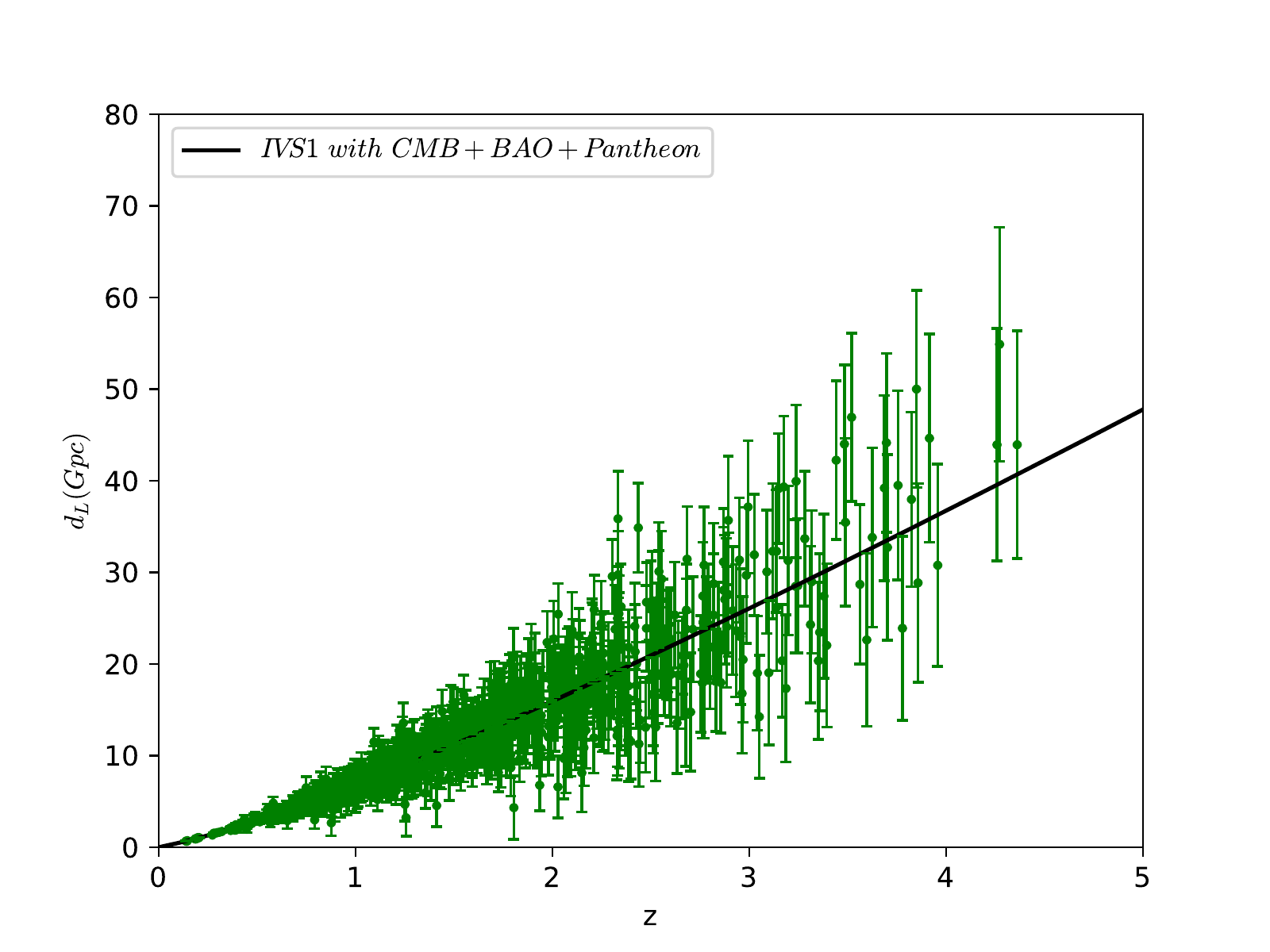}
    \includegraphics[width=0.43\textwidth]{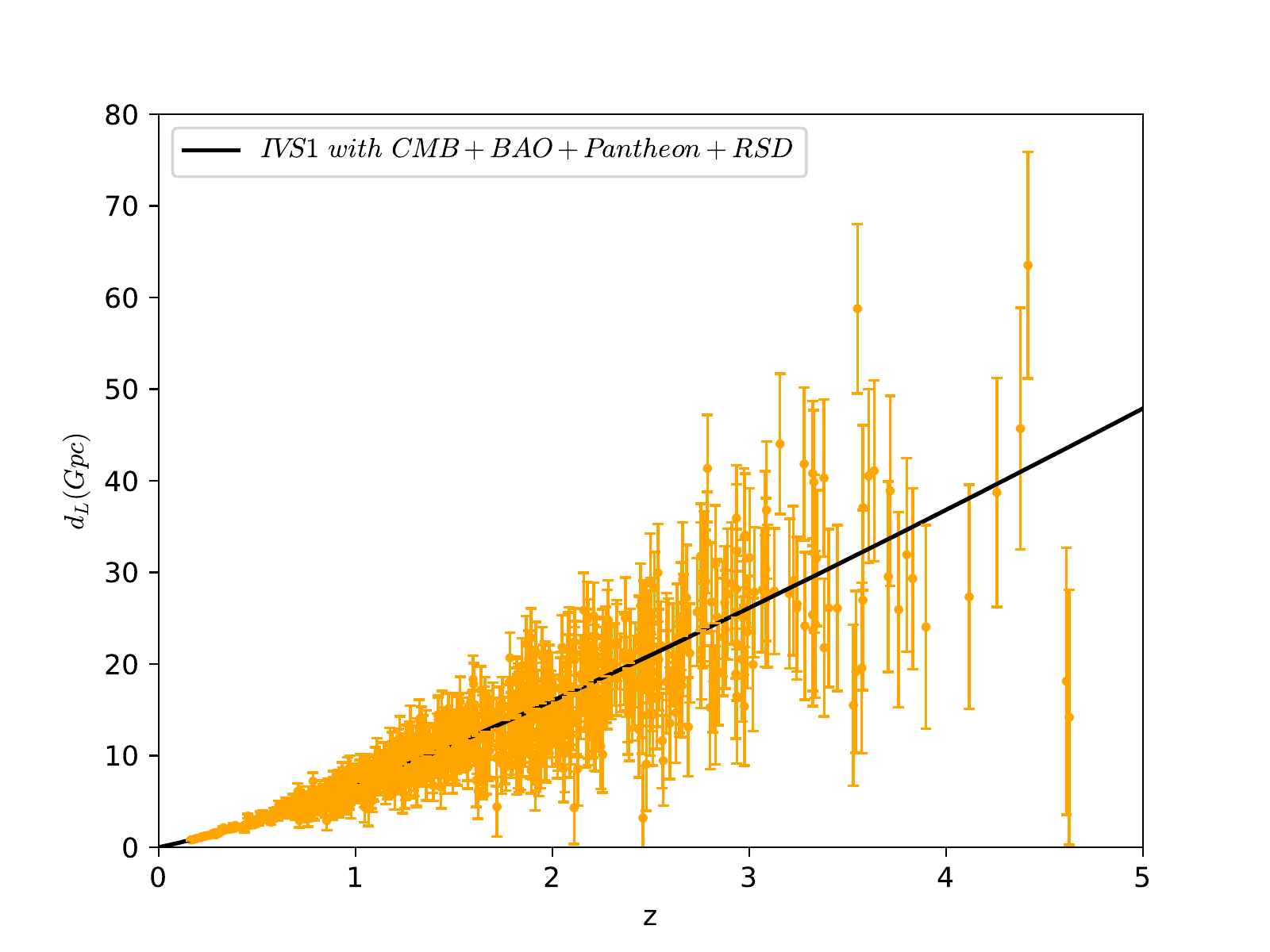}
    \includegraphics[width=0.43\textwidth]{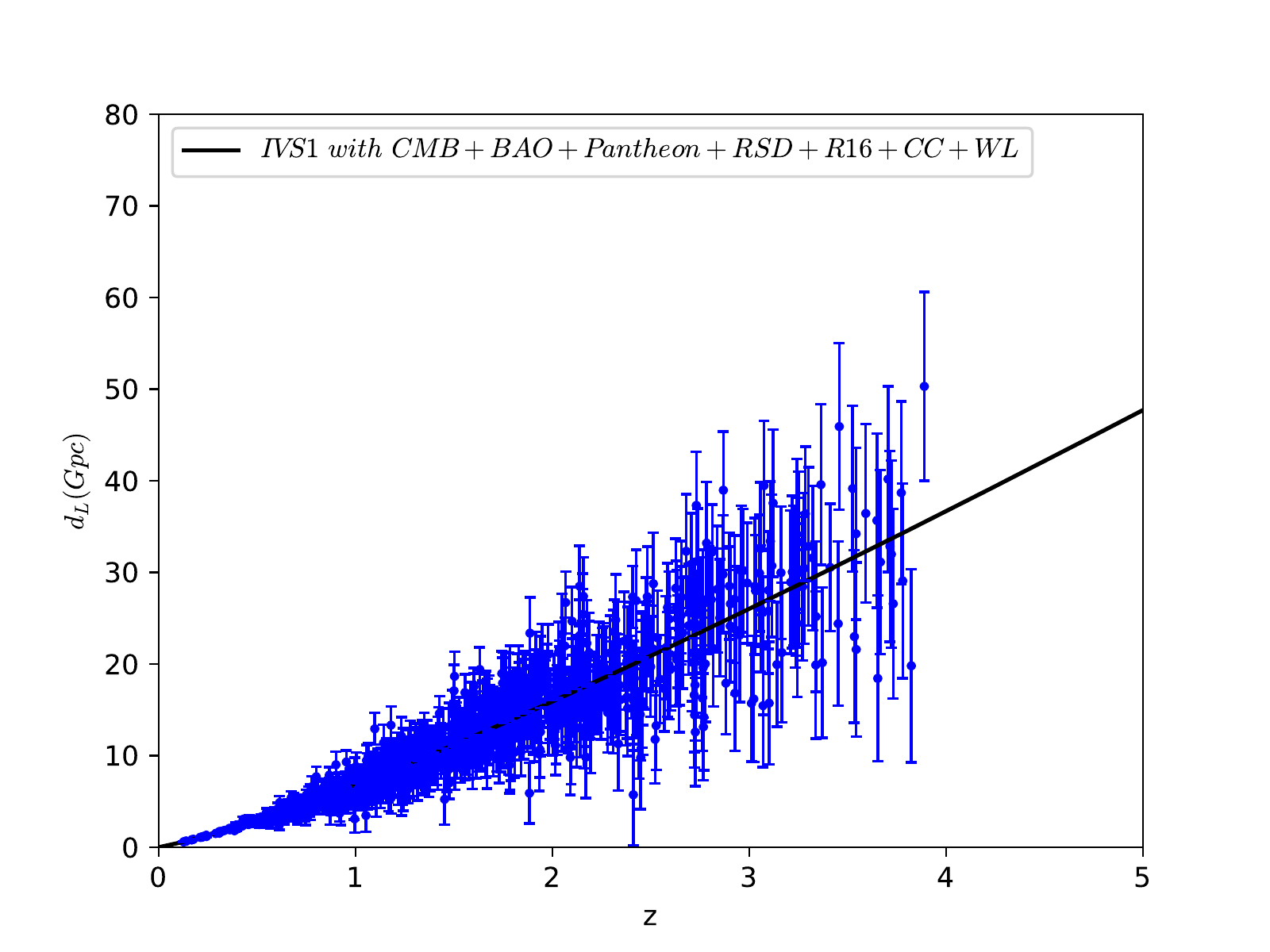}
    \caption{The description of the figure is as follows. Here we have considered the IVS1 model as the fiducial one. We first constrain the free and derived  parameters of IVS1 using all the datasets, namely, CMB, CMB+BAO, CMB+BAO+Pantheon and CMB+BAO+Pantheon+RSD and CMB+BAO+Pantheon+RSD+R16+CC+WL. After that we use the best-fit values of all the constrained 
    parameters for each dataset (CMB, CMB+BAO and others) to generate the corresponding GW catalogue containing 1000 simulated GW events. In each plot we show $d_L (z)$ vs $z$ catalogue with corresponding error bars for 1000 simulated GW events. The upper left graph is for CMB alone, upper right graph is for CMB+BAO, the lower left graph is for CMB+BAO+Pantheon, lower right graph is for CMB+BAO+Pantheon+RSD and the bottom graph is for the dataset CMB+BAO+Pantheon+RSD+R16+CC+WL. }
    \label{fig:gw_error_ivs1}
\end{figure*}
\begin{figure*}
    \centering
    \includegraphics[width=0.43\textwidth]{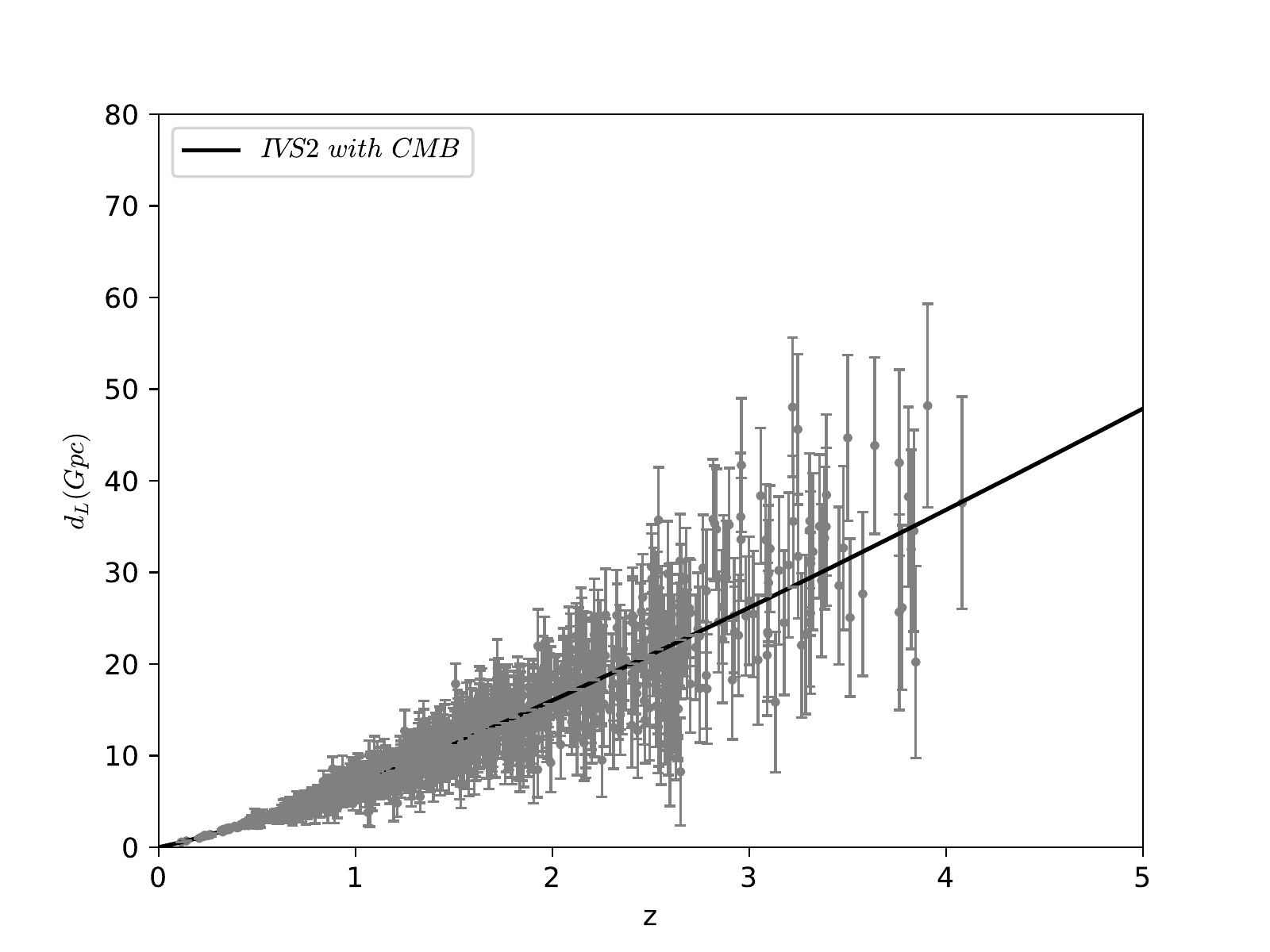}
    \includegraphics[width=0.43\textwidth]{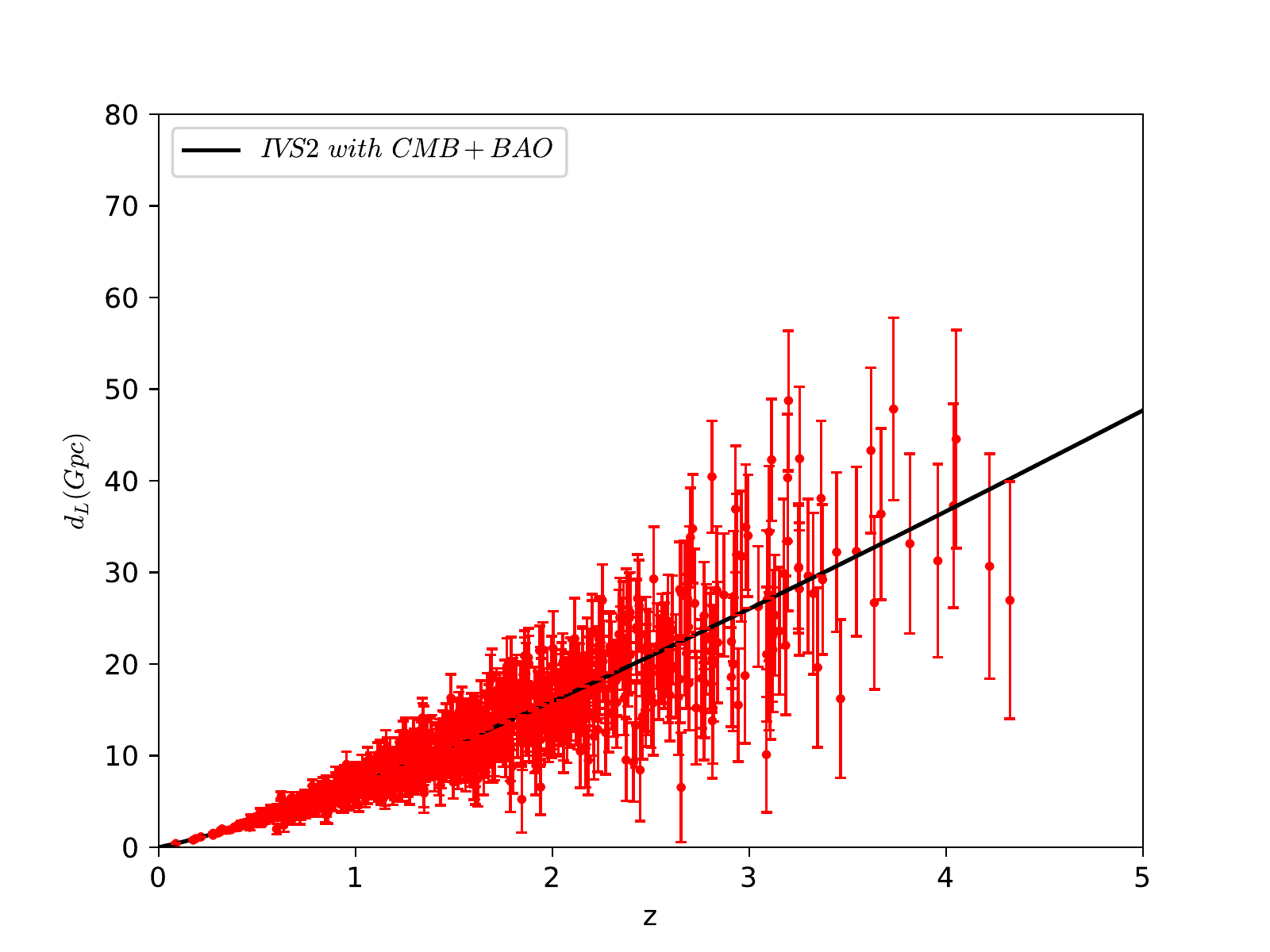}
    \includegraphics[width=0.43\textwidth]{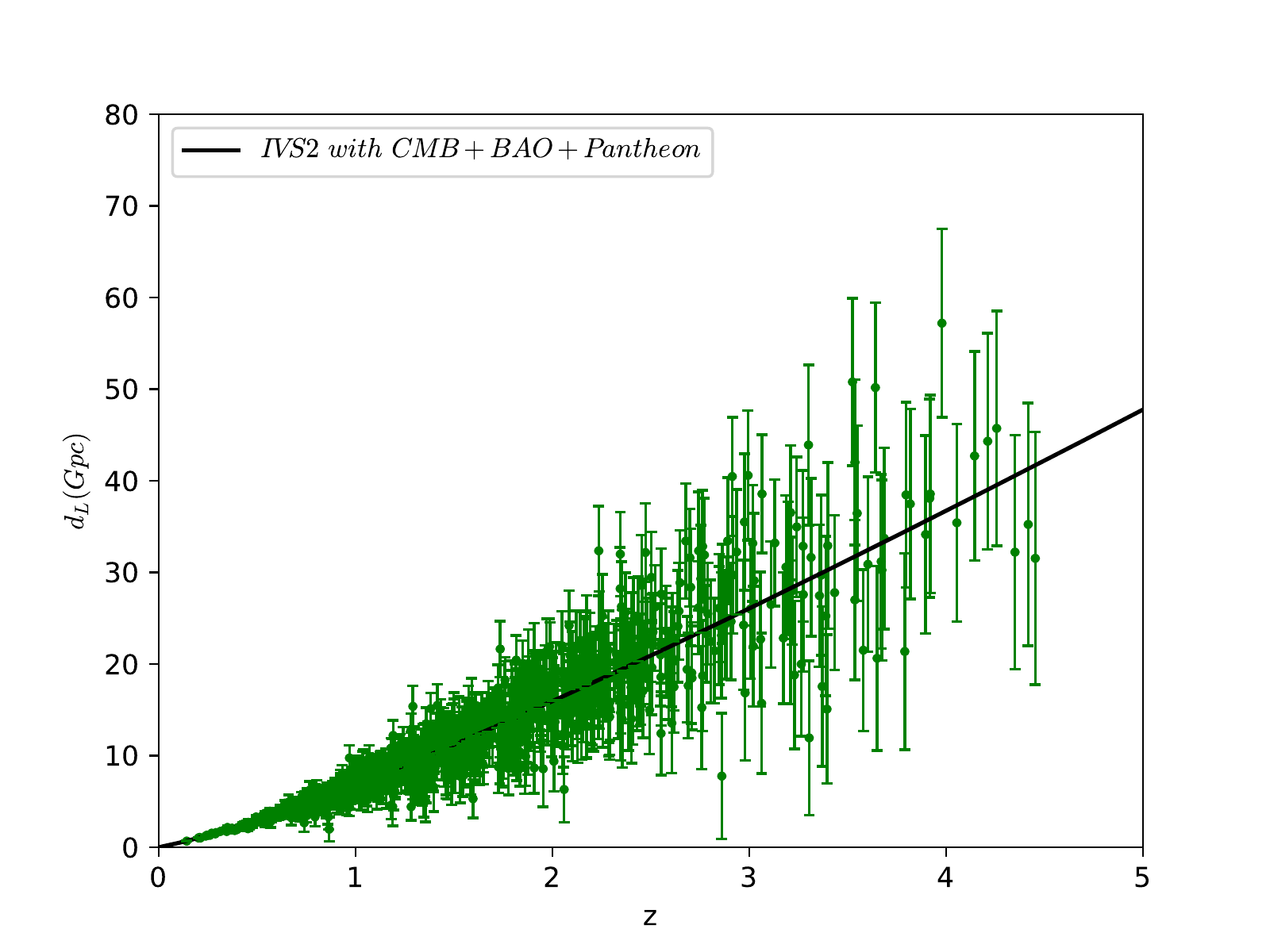}
    \includegraphics[width=0.43\textwidth]{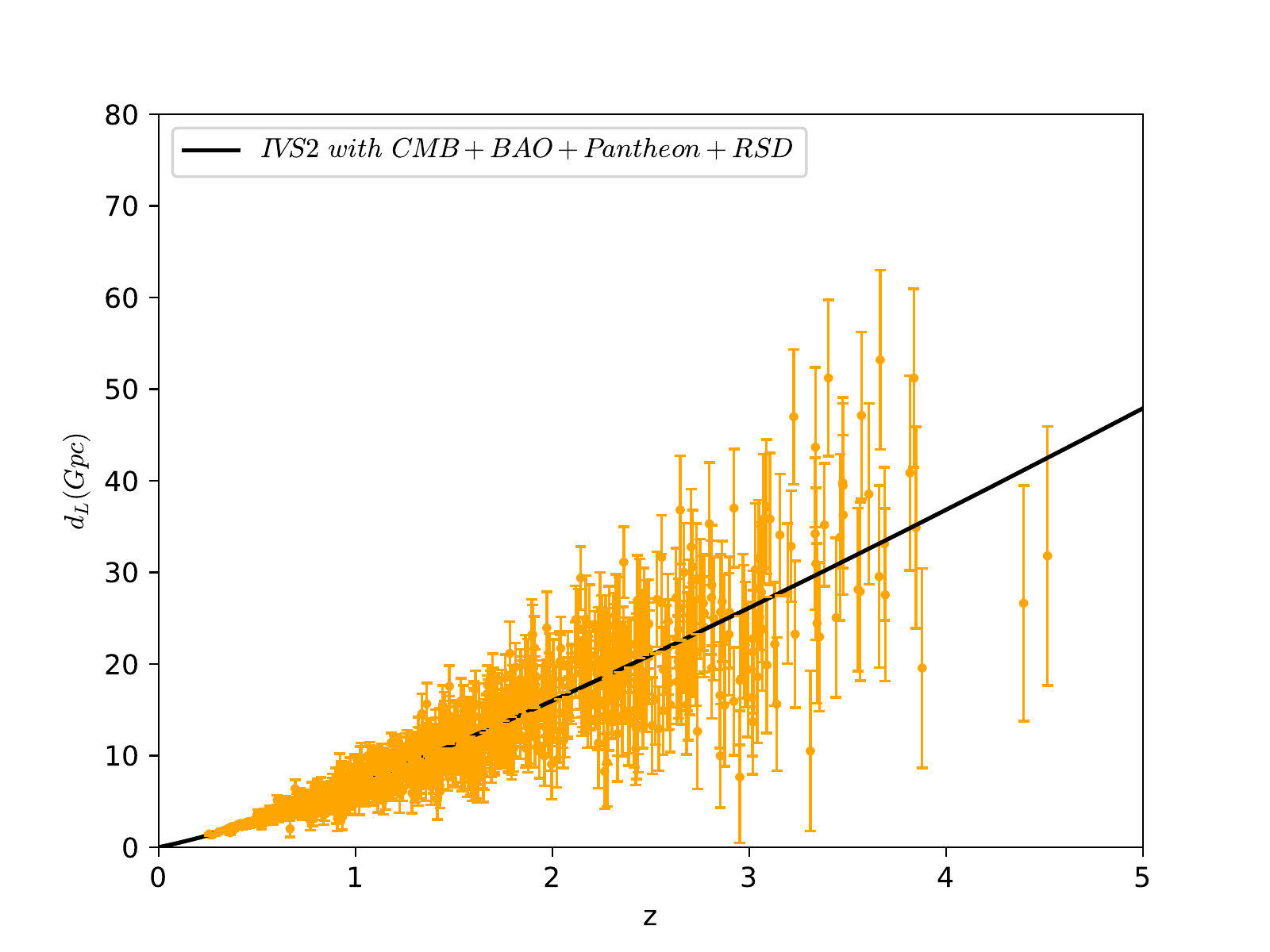}
    \includegraphics[width=0.43\textwidth]{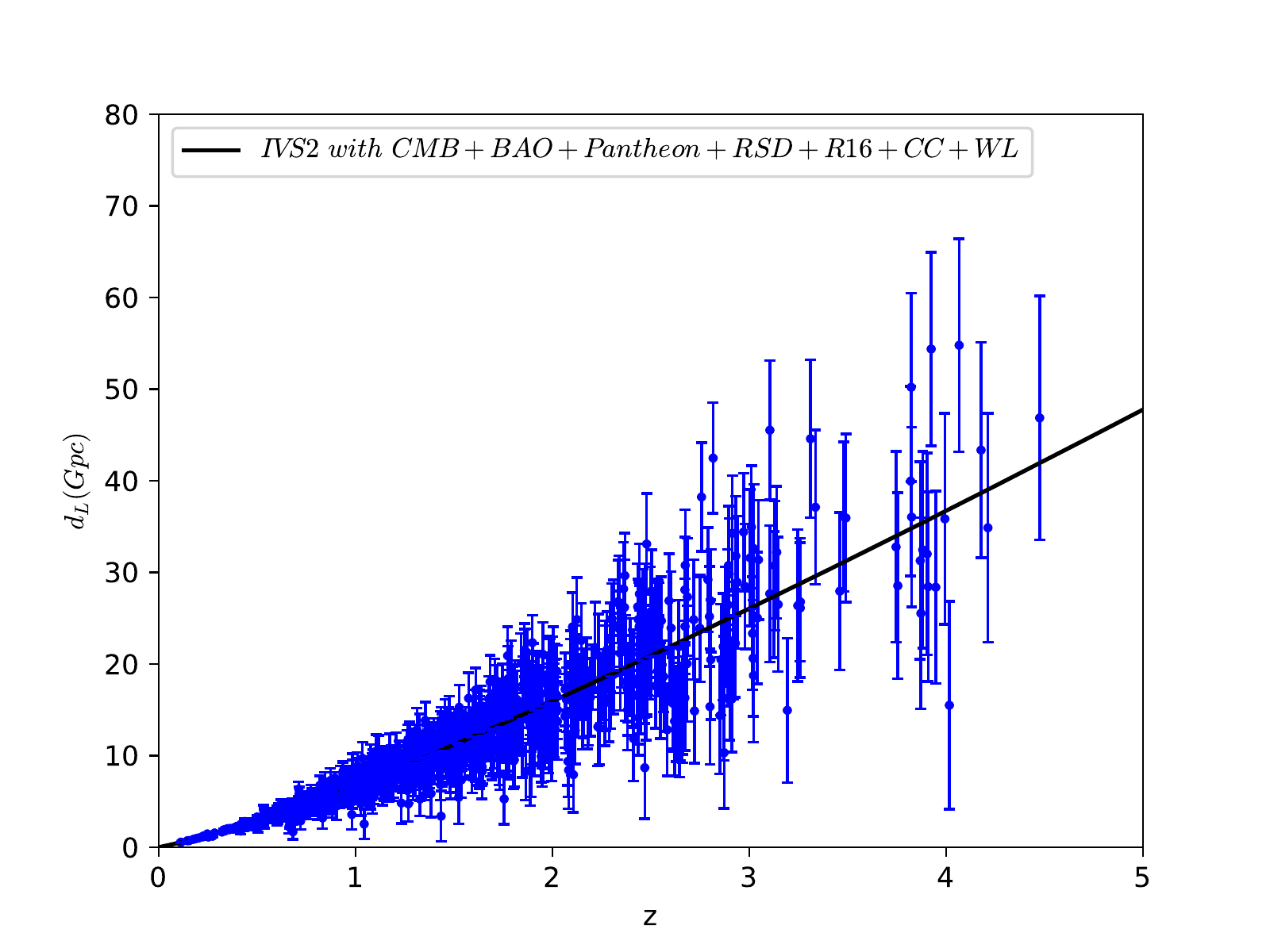}
    \caption{The description of the figure is as follows. Here we have considered the IVS2 model as the fiducial one. We first constrain the free and derived  parameters of IVS2 using all the datasets, namely, CMB, CMB+BAO, CMB+BAO+Pantheon and CMB+BAO+Pantheon+RSD and CMB+BAO+Pantheon+RSD+R16+CC+WL, and after that we use the best-fit values of all the constrained parameters for each dataset (CMB, CMB+BAO and others) to generate the corresponding GW catalogue containing 1000 simulated GW events. In each plot we show $d_L (z)$ vs $z$ catalogue with corresponding error bars for 1000 simulated GW events. The upper left graph is for CMB alone, upper right graph is for CMB+BAO, the lower left graph is for CMB+BAO+Pantheon, lower right graph is for CMB+BAO+Pantheon+RSD and the bottom graph is for the dataset CMB+BAO+Pantheon+RSD+R16+CC+WL.}
    \label{fig:gw_error_ivs2}
\end{figure*}

With the above descriptions, we are now at the final stage to use the GWSS data directly into our analysis. 
The analysis with GW data is same compared to the usual cosmological data.  For GWSS measurements comprising $N$ simulated data points, the $\chi^2$ function is,
\begin{align}
\chi_{\rm GW}^2=\sum\limits_{i=1}^{N}\left[\frac{\bar{d}_L^i-d_L(\bar{z}_i;\vec{\Theta})}{\bar{\sigma}_{d_L}^i}\right]^2,
\label{equa:chi2}
\end{align}
in which as already mentioned above, $\bar{d}_L^i$, $\bar{\sigma}_{d_L}^i$ are respectively the luminosity distance and the corresponding error at a particular redshift $z_i$. Let us note that by $\vec{\Theta}$ we represent the cosmological parameters involved with the models.

The likelihood analysis of this work follows $\mathcal{L} \propto e^{-\sum \chi^2_i/2}$, where $\chi^2_i$ represents the corresponding $\chi^2$ for the $i$-th dataset employed in this work. So, we clarify that when we use only the standard cosmological datasets, then $i$ must belong to CMB, BAO, RSD etc (see subsection \ref{subsec-data}), while if we want to include GWs with the standard datasets, then $\chi^2$ should include both from standard datasets and simulated GW catalogue.
The global fitting has been done by using \texttt{CosmoMC} \cite{Lewis:2002ah, Lewis:1999bs}, a publicly available Monte-Carlo Markov Chain package which  (i) is equipped with a convergence diagnostic based on the Gelman and Rubin statistic, (ii) implements an efficient sampling of the posterior distribution using the fast/slow parameter decorrelations \cite{Lewis:2013hha}, and (iii) includes the support for the Planck data release 2015 Likelihood code \cite{Aghanim:2015xee} (see the freely available code at \url{http://cosmologist.info/cosmomc/}). We note that in Table \ref{tab:priors} we have shown the priors imposed on all free parameters for the purpose of statistical analysis. The readers may wonder why we use the Planck 2015 likelihood data  \cite{Aghanim:2015xee}, as the Planck 2018 cosmological parameters are now already available \cite{Aghanim:2018eyx}. The reason is that Planck 2018 likelihood code is still not made public. However, after the new Planck 2018 likelihood code is available it will be interesting to redo the same analysis in order to see how the new data affect the cosmological parameters compared to their constraints obtained from Planck 2015.

\begin{figure*}
\includegraphics[width=0.8\textwidth]{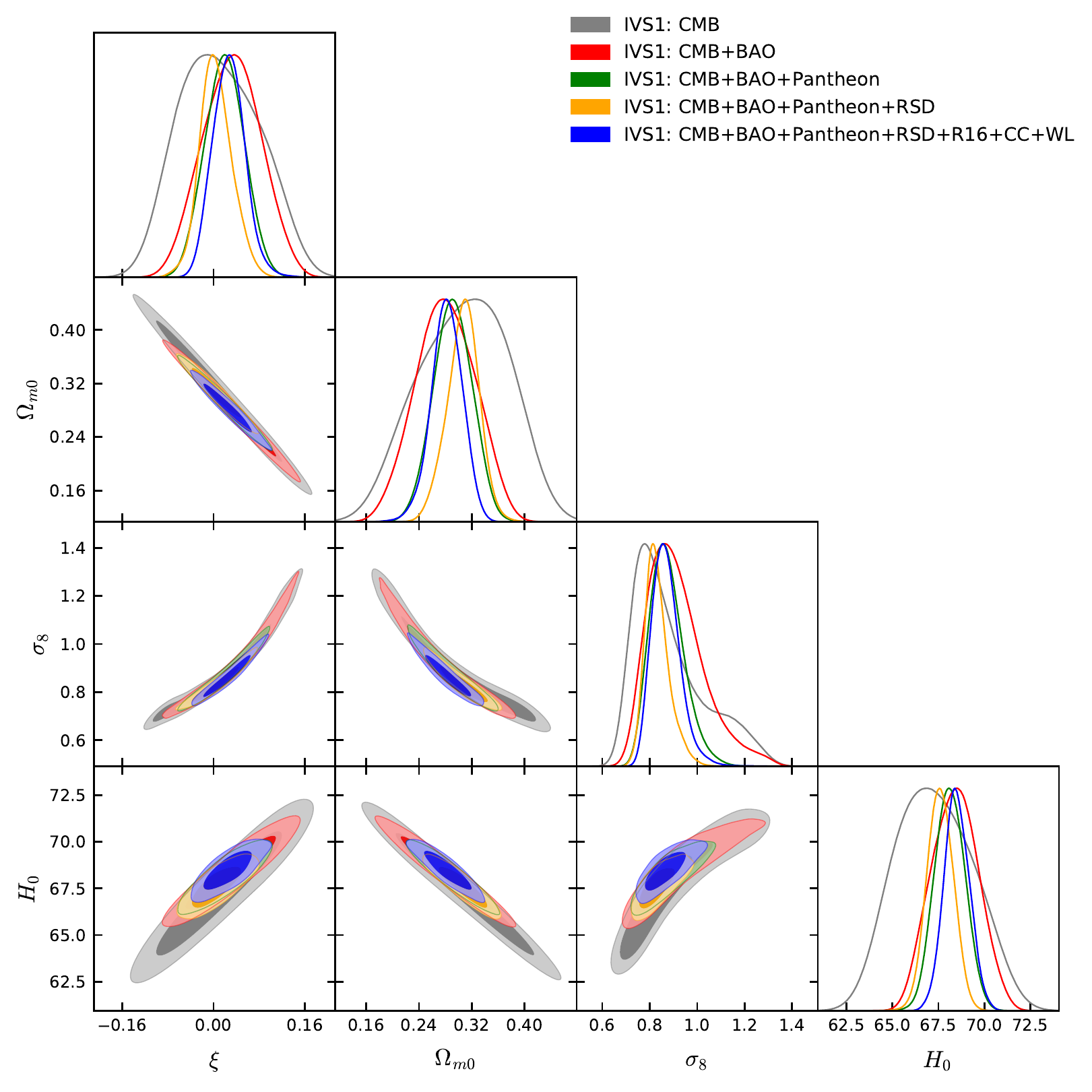}
\caption{The figure for IVS1 shows two different things as follows. It contains the 68\% and 95\% confidence level contour plots among several combinations of the model parameters using only the standard cosmological probes namely CMB, BAO, Pantheon, RSD, R16, CC and WL. Also, it shows the one dimensional marginalized posterior distributions of some selected parameters. We note that here $\Omega_{m0}$ is the present value of $\Omega_m = \Omega_c +\Omega_b$, and $H_0$ is in the units of Km/sec/Mpc. }
\label{fig:real_IVS1}
\end{figure*}

\begingroup                               
\squeezetable                              
\begin{center}                              
\begin{table*}
\begin{tabular}{ccccccccc}                      
\hline\hline   
Parameters & CMB & CB & CBP & CBPR & CBPRR16CW\\ \hline
$\Omega_c h^2$ & $ 0.107_{-    0.023-    0.039}^{+    0.027+    0.037}$ & $    0.108_{-    0.015- 0.033}^{+    0.019+    0.032}$ & $    0.112_{- 0.010-    0.021}^{+0.012+    0.021}$ & $    0.1171_{-    0.0071-  0.018}^{+    0.0097+    0.016}$ &  $ 0.1092_{-    0.0082-    0.018}^{+    0.0090+    0.018}$ \\

$\Omega_b h^2$ & $ 0.02220_{-    0.00016-    0.00030}^{+    0.00016+    0.00031}$ & $    0.02226_{-    0.00015-    0.00029}^{+    0.00015+    0.00030}$ & $    0.02227_{-    0.00015-    0.00028}^{+    0.00015+    0.00029}$ & $    0.02226_{-    0.00015-    0.00029}^{+    0.00015+    0.00030}$ & $    0.02230_{-    0.00014-    0.00028}^{+    0.00015+    0.00028}$ \\

$100\theta_{MC}$ & $ 1.0412_{-    0.0017-    0.0024}^{+    0.0013+    0.0026}$ &  $    1.0412_{-    0.0011-    0.0019}^{+    0.0009+    0.0020}$ & $    1.04096_{-    0.00068-    0.0012}^{+    0.00061+    0.0013}$ & $    1.04065_{-    0.00055-    0.0010}^{+    0.00049+    0.0011}$ & $    1.04113_{-    0.00059-    0.0011}^{+    0.00051+    0.0011}$  \\

$\tau$ & $ 0.078_{-    0.017-    0.033}^{+    0.017+    0.033}$ & $    0.084_{-    0.017-    0.034}^{+    0.017+    0.033}$ & $    0.085_{-    0.017-    0.033}^{+    0.017+    0.031}$ & $    0.074_{-    0.016-    0.031}^{+    0.017+    0.032}$ & $    0.067_{-    0.016-    0.034}^{+    0.018+    0.031}$\\

$n_s$ & $ 0.9734_{-    0.0045-    0.0090}^{+    0.0044+    0.0089}$ & $    0.9755_{-    0.0045-    0.0082}^{+    0.0041+    0.0084}$ & $    0.9759_{-    0.0040-    0.0080}^{+    0.0041+    0.0078}$ & $    0.9752_{-    0.0042-    0.0080}^{+    0.0042+    0.0079}$ & $    0.9762_{-    0.0041-    0.0077}^{+    0.0040+    0.0083}$ \\

${\rm{ln}}(10^{10} A_s)$ & $ 3.100_{-    0.033-    0.064}^{+    0.033+    0.065}$ & $    3.109_{-    0.034-    0.066}^{+    0.034+    0.065}$ & $    3.111_{-    0.033-    0.066}^{+    0.034+    0.061}$ & $    3.089_{-    0.032-    0.061}^{+    0.032+    0.063}$ & $    3.074_{-    0.032-    0.066}^{+    0.034+    0.061}$ \\

$\xi$ & $0.034_{-    0.068-    0.13}^{+    0.083+    0.12}$ & $ 0.032_{-    0.054-    0.098}^{+    0.051+    0.098}$ & $    0.020_{-    0.035-    0.065}^{+    0.034+    0.066}$ & $    0.005_{-    0.032-    0.055}^{+    0.025+    0.060}$ & $    0.027_{-    0.029-    0.057}^{+    0.027+    0.056}$ \\

$\Omega_{m0}$ & $ 0.285_{-    0.082-    0.11}^{+    0.062+    0.12}$ & $    0.280_{-    0.045-    0.089}^{+    0.047+ 0.086}$ & $    0.291_{-    0.029- 0.057}^{+    0.030+    0.058}$  & $    0.307_{-    0.022-    0.051}^{+    0.027+    0.046}$ & $    0.282_{-    0.023-    0.049}^{+    0.024+    0.048}$ \\

$\sigma_8$ & $ 0.93_{-    0.22-    0.26}^{+    0.10+    0.33}$ & $ 0.91_{-    0.15-    0.22}^{+    0.08+    0.26}$ & $    0.873_{-    0.085-    0.14}^{+    0.059+    0.15}$  & $    0.832_{-    0.060-    0.097}^{+    0.038+    0.12}$ & $    0.869_{-    0.069-    0.12}^{+    0.048+    0.12}$ \\

$H_0$ & $ 68.0_{-    1.9-    3.8}^{+    2.5+    3.5}$ & $ 68.4_{-    1.3-    2.4}^{+    1.3+    2.5}$ & $   68.12_{-    0.83-    1.6}^{+    0.82+    1.6}$ & $   67.60_{-    0.70-    1.4}^{+    0.72+    1.4}$ & $   68.50_{-    0.67-    1.3}^{+    0.70+    1.4}$ \\
\hline\hline                                                                                                                    
\end{tabular} 
\caption{Observational constraints on the first interaction model IVS1 of (\ref{ivs1}) have been shown using the standard cosmological probes CMB, BAO, Pantheon, RSD, R16, CC and WL. The parameter $\Omega_{m0}$ is the present value of the matter density parameter $\Omega_m = \Omega_c+\Omega_b$, and $H_0$ is in the units of km/sec/Mpc. Here, CB = CMB+BAO, CBP = CMB+BAO+Pantheon, CBPR = CMB+BAO+Pantheon+RSD, CBPRR16CW = CMB+BAO+Pantheon+RSD+R16+CC+WL.}
\label{tab:ivs1-noGW}                                                                                                   
\end{table*}                                                                                                                     
\end{center}                                                                                                                    
\endgroup 
\begingroup                              
\squeezetable  
\begin{center} 
\begin{table*}                                                                                                                   
\begin{tabular}{ccccccccccccccc}                                    \hline\hline
Parameters & CG & CBG & CBPG & CBPRG & CBPRR16CWG \\ \hline
$\Omega_c h^2$ & $0.109_{-    0.010-    0.022}^{+    0.012+    0.019}$ & $    0.118_{-    0.010-    0.020}^{+    0.011+    0.019}$ & $    0.1170_{-    0.0092-    0.016}^{+    0.0090+    0.018}$ & $    0.1170_{-    0.0092-    0.016}^{+    0.0090+    0.018}$ & $    0.1136_{-    0.0060-    0.012}^{+    0.0058+    0.012}$\\

$\Omega_b h^2$ & $    0.02218_{-    0.00014-    0.00028}^{+    0.00015+    0.00029}$  &  $    0.02204_{-    0.00016-    0.00028}^{+    0.00014 +    0.00029}$  & $    0.02222_{-    0.00014-    0.00028}^{+    0.00015+    0.00028}$ & $    0.02228_{-    0.00014-    0.00027}^{+    0.00014+    0.00029}$ & $    0.02238_{-    0.00013-    0.00028}^{+    0.00014+    0.00028}$ \\

$100\theta_{MC}$ & $    1.04107_{-    0.00067-    0.0011}^{+    0.00054+    0.0013}$ & $    1.04031_{-    0.00060-    0.0012}^{+    0.00060+    0.0012}$ & $    1.04065_{-    0.00048-    0.00093}^{+    0.00047+    0.00097}$ &  $    1.04069_{-    0.00048-    0.00098}^{+    0.00047+    0.00098}$ & $    1.04099_{-    0.00036-    0.00075}^{+    0.00036+    0.00072}$\\

$\tau$ & $    0.078_{- 0.017-    0.034}^{+    0.017+    0.033}$ & 
$ 0.074_{- 0.017-    0.032}^{+ 0.017+    0.034}$ & $    0.082_{-    0.017-    0.033}^{+    0.017+    0.033}$ & $    0.075_{-    0.017-    0.034}^{+    0.017+    0.033}$ & $    0.074_{-    0.017-    0.034}^{+    0.017+    0.033}$ \\

$n_s$ & $    0.9728_{- 0.0044-    0.0082}^{+    0.0042+    0.0088}$ & $    0.9712_{-  0.0040-    0.0081}^{+    0.0041+    0.0083}$ & $    0.9749_{- 0.0041-    0.0074}^{+    0.0040+    0.0078}$ &  $    0.9755_{- 0.0043-    0.0082}^{+    0.0042+    0.0087}$ & $    0.9781_{- 0.0040-    0.0077}^{+    0.0040+    0.0079}$ \\

${\rm{ln}}(10^{10} A_s)$ & $    3.100_{-    0.032-    0.066}^{+    0.035+    0.063}$ & $    3.093_{-    0.033-    0.063}^{+    0.034+    0.066}$ &  $    3.106_{-    0.034-    0.066}^{+    0.034+    0.065}$ & $    3.090_{-    0.032-    0.064}^{+    0.033+    0.062}$ & $    3.086_{-    0.033-    0.066}^{+    0.034+    0.064}$ \\

$\xi$ & $    0.034_{-    0.038-    0.068}^{+    0.035+    0.070}$  & $    0.0056_{-    0.036-    0.067}^{+    0.037+    0.069}$ & $    0.0070_{-    0.029-    0.054}^{+    0.028+    0.056}$ & $    0.0045_{-    0.031-    0.063}^{+    0.033+    0.056}$ & $    0.012_{-    0.020-    0.041}^{+    0.021+    0.039}$ \\

$\Omega_{m0}$ & $    0.285_{-    0.028-    0.056}^{+    0.028+    0.053}$ & $    0.317_{-    0.029-    0.055}^{+    0.030+    0.052}$  & $    0.306_{-    0.022-    0.044}^{+    0.022+    0.043}$ & $    0.306_{-    0.026-    0.043}^{+    0.023+    0.046}$ & $    0.292_{-    0.015-    0.029}^{+    0.014+    0.030}$ \\

$\sigma_8$ & $    0.901_{-    0.094-    0.15}^{+    0.060+    0.17}$ & $    0.841_{-    0.077-    0.12}^{+    0.055+    0.13}$ &$    0.844_{-    0.061-    0.10}^{+    0.046+    0.11}$ & $    0.832_{-    0.062-    0.11}^{+    0.049+    0.11}$ & $    0.839_{-    0.042-    0.073}^{+    0.037+    0.081}$ \\

$H_0$ & $   67.94_{-    0.64-    1.3}^{+    0.65+    1.3}$ & $   66.77_{-    0.65-    1.2}^{+    0.66+    1.3}$ & $   67.52_{-    0.56-    1.0}^{+    0.51+    1.0}$ & $   67.69_{-    0.57-    1.0}^{+    0.56+    1.0}$ & $   68.42_{-    0.34-    0.66}^{+    0.33+    0.67}$ \\
\hline\hline                                                                                                                    
\end{tabular}
\caption{Observational constraints on the first interaction model IVS1 of (\ref{ivs1}) have been shown after the inclusion of the gravitational waves data with the standard cosmological probes. The parameter $\Omega_{m0}$ is the present value of the matter density parameter $\Omega_m = \Omega_c+\Omega_b$, and $H_0$ is in the units of km/sec/Mpc. Here, CG = CMB+GW, CBG = CMB+BAO+GW, CBPG = CMB+BAO+Pantheon+GW, CBPRG = CBPR = CMB+BAO+Pantheon+RSD+GW, CBPRR16CWG = CMB+BAO+Pantheon+RSD+R16+CC+WL+GW.} \label{tab:ivs1-withGW}                                                                                                   
\end{table*}                                                                                                                     
\end{center}                                                                                                                    
\endgroup
\begin{figure*}
\includegraphics[width=1.0\textwidth]{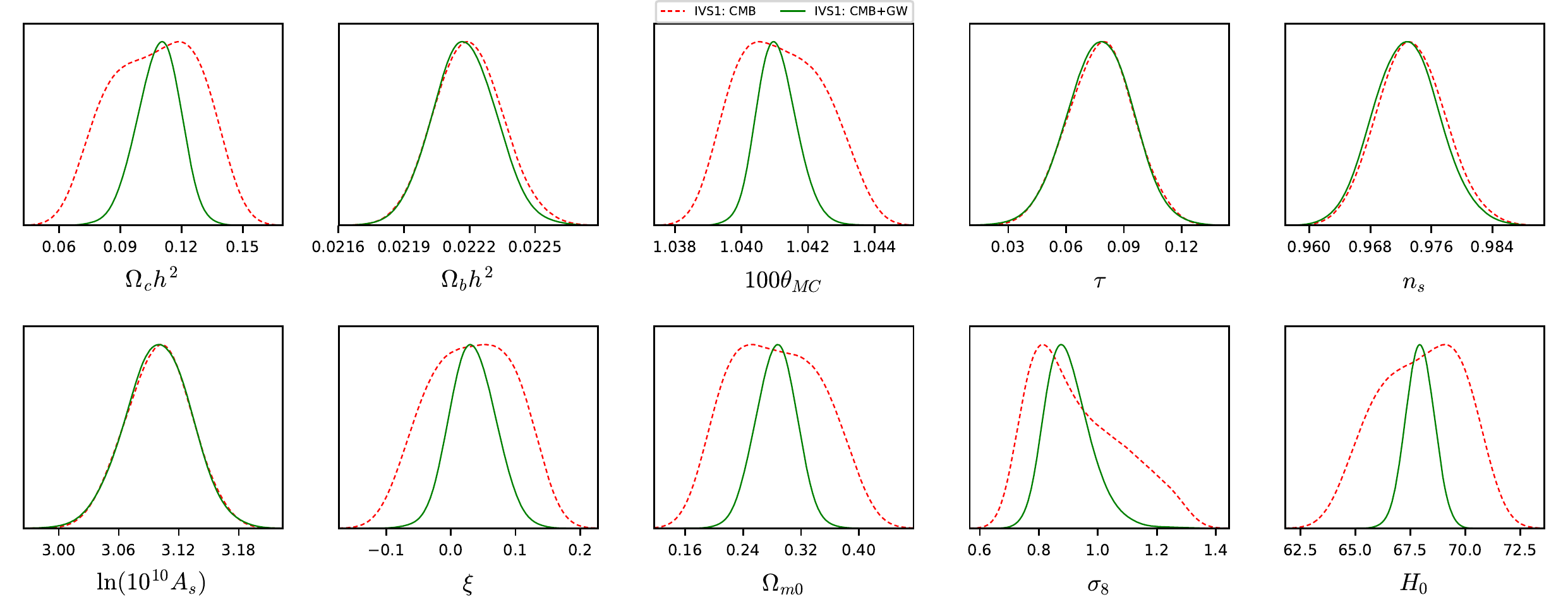}
\caption{We compare the one dimensional marginalized posterior distributions for some selected parameters of the interacting scenario IVS1 considering the observational datasets CMB and CMB+GW. }
\label{fig:1DIVS1-CG}
\end{figure*}
\begin{figure*}
\includegraphics[width=0.8\textwidth]{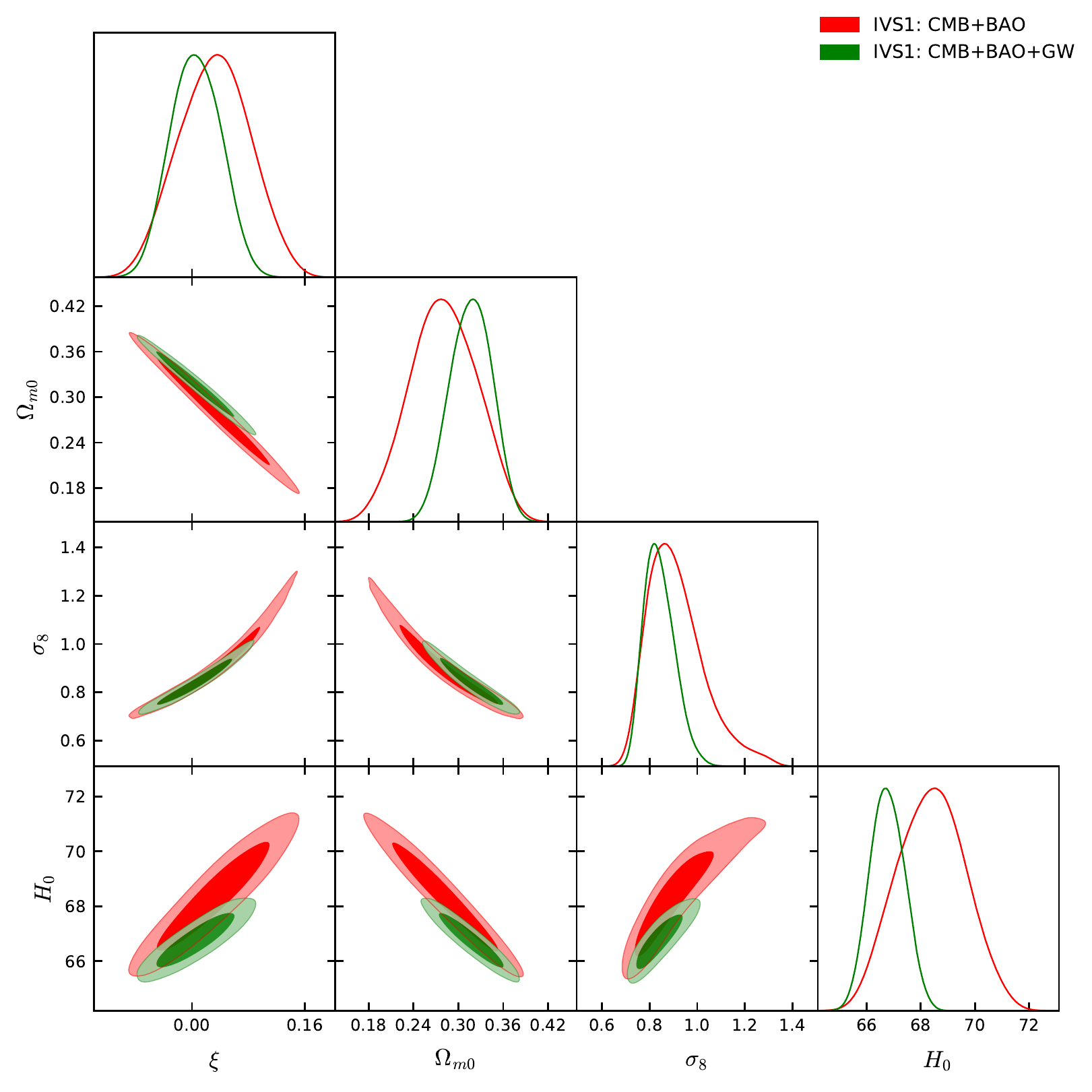}
\caption{We compare the cosmological constraints on the free parameters of IVS1 for the observational datasets CMB+BAO and CMB+BAO+GW. }
\label{fig:triIVS1-CBG}
\end{figure*}
\begin{figure*}
\includegraphics[width=0.8\textwidth]{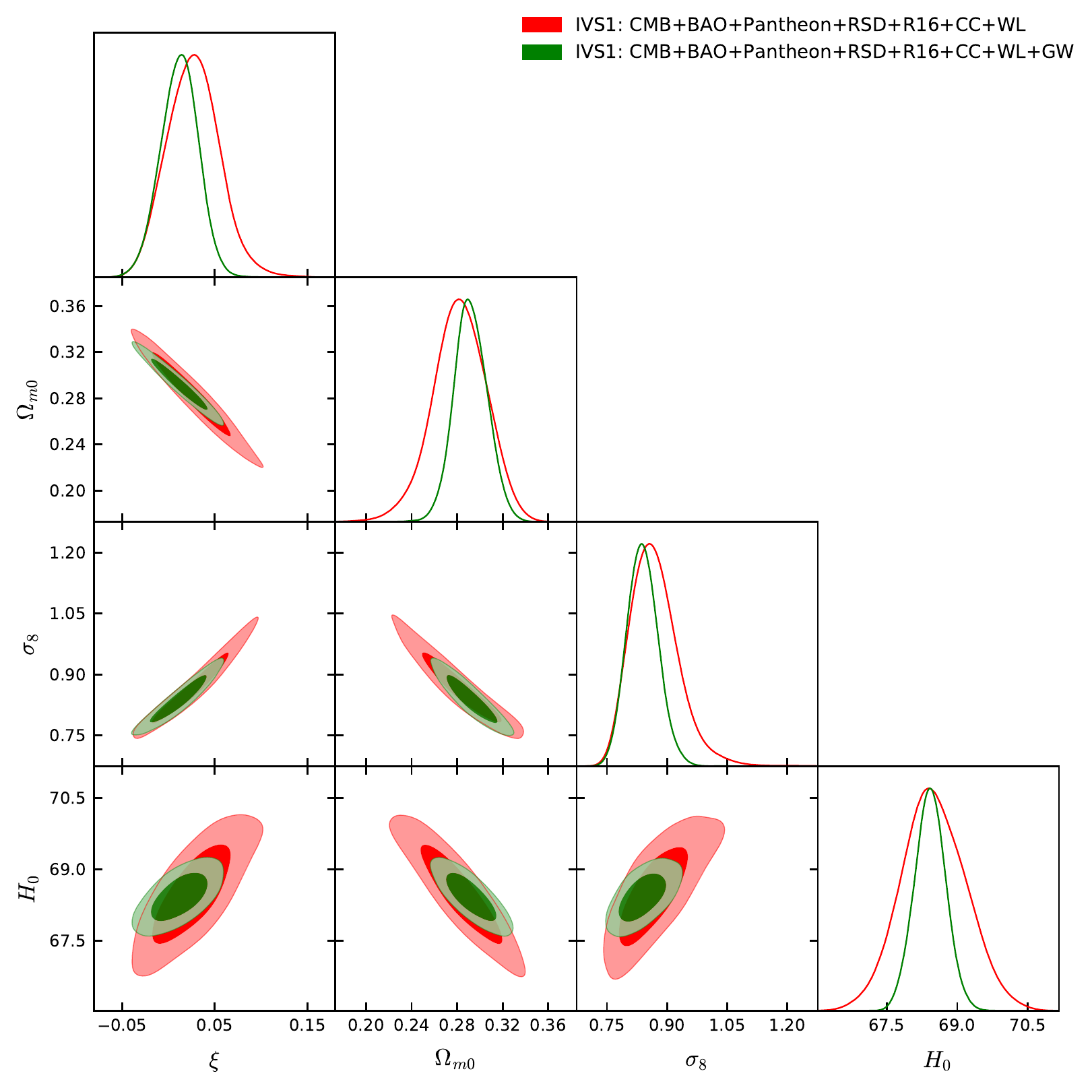}
\caption{We compare the cosmological constraints on the free parameters of IVS1 for the observational datasets CMB+BAO+Pantheon+RSD+R16+CC+WL and CMB+BAO+Pantheon+RSD+R16+CC+WL+GW. }
\label{fig:tri_IVS1-full}
\end{figure*}                                                                                                

\section{Results}
\label{sec-results}

In this section we  describe the main  results that we find from the interacting models considered in this work. For both the interaction models we have considered exactly similar observational datasets, namely, CMB, CMB+BAO, CMB+BAO+Pantheon, CMB+BAO+Pantheon+RSD, CMB+BAO+Pantheon+RSD+R16+CC+WL, to be homogeneous throughout the description.  Before presenting all the results for IVS1 and IVS2 we comment on the fiducial model. In order to generate the GW catalogue we have assumed IVS1 and IVS2 as the fiducial models. When we consider IVS1 as the fiducial model, we first consider the best-fit values for all the free and derived parameters of this model extracted from any observational dataset. Then we use these best-fit values to generate the corresponding GW catalogue that contains 1000 simulated GW events. In Fig.  \ref{fig:gw_error_ivs1} we have shown the relation between $d_L (z)$ vs. $z$ (with error bars on $d_L (z)$) for 1000 GW simulated GW events. We then use this catalogue as the forecasted dataset with the standard cosmological probes using the $\chi^2$-technique described at the end of section \ref{sec-data}. 
In a similar fashion we consider IVS2 as the fiducial model and do the same thing similar to IVS1. Fig. \ref{fig:gw_error_ivs2} shows the catalogue for 1000 simulated GW events for IVS2. In what follows we describe the results that we extract from IVS1 and IVS2 after the inclusion of the simulated GW data.

\begingroup 
\squeezetable                                                   
\begin{center}                                 
\begin{table*}
\begin{tabular}{ccccccccccccc}                                  \hline\hline                                
Parameters & CMB & CB & CBP & CBPR & CBPRR16CW \\ \hline
$\Omega_c h^2$ & $    0.118_{-    0.033-    0.061}^{+    0.037+    0.060}$ & $    0.112_{-    0.011-    0.025}^{+    0.014+    0.022}$ & $    0.1132_{-    0.0058-    0.016}^{+    0.0078+    0.013}$ & $    0.1190_{-    0.0060-    0.013}^{+    0.0062+    0.012}$ & $    0.1122_{-    0.0056-    0.011}^{+    0.0061+    0.011}$ \\

$\Omega_b h^2$ & $    0.02220_{-    0.00015-    0.00029}^{+    0.00015+    0.00031}$ & $    0.02226_{-    0.00015-    0.00031}^{+    0.00015+    0.00030}$ & $    0.02226_{-    0.00016-    0.00028}^{+    0.00014+    0.00029}$ & $    0.02226_{-    0.00015-    0.00031}^{+    0.00015+    0.00030}$ & $    0.02229_{-    0.00015-    0.00029}^{+    0.00015+    0.00029}$  \\

$100\theta_{MC}$ & $    1.0406_{-    0.0023-    0.0031}^{+    0.0016+    0.0035}$ & $    1.04094_{-    0.00077-    0.0013}^{+    0.00063+    0.0014}$ & $    1.04090_{-    0.00051-    0.00085}^{+    0.00041+    0.00096}$ & $    1.04055_{-    0.00041-    0.00083}^{+    0.00042+    0.00084}$ & $    1.04097_{-    0.00039-    0.00076}^{+    0.00040+    0.00078}$ \\

$\tau$ & $    0.079_{-    0.017-    0.034}^{+    0.017+    0.033}$ & $    0.085_{-    0.016-    0.033}^{+    0.016+    0.032}$ & $    0.086_{-    0.016-    0.034}^{+    0.019+    0.031}$ &  $    0.075_{-    0.017-    0.034}^{+    0.017+    0.034}$ & $    0.068_{-    0.016-    0.034}^{+    0.016+    0.032}$ \\

$n_s$ & $    0.9733_{-    0.0047-    0.0085}^{+    0.0041+    0.0089}$ & $    0.9754_{-    0.0045-    0.0084}^{+    0.0043+    0.0087}$ & $    0.9758_{-    0.0041-    0.0083}^{+    0.0043+    0.0080}$ & $    0.9754_{-    0.0040-    0.0081}^{+    0.0041+    0.0083}$ & $    0.9759_{-    0.0045-    0.0081}^{+    0.0041+    0.0084}$ \\

${\rm{ln}}(10^{10} A_s)$ & $    3.101_{-    0.033-    0.065}^{+    0.033+    0.064}$ & $    3.111_{-    0.031-    0.064}^{+    0.031+    0.062}$ & $    3.113_{-    0.032-    0.065}^{+    0.037+    0.061}$ & $    3.090_{-    0.033-    0.067}^{+    0.033+    0.065}$ & $    3.076_{-    0.031-    0.065}^{+    0.032+    0.061}$ \\

$\xi$ & $    0.02_{-    0.27-    0.42}^{+    0.22+    0.42}$ & $    0.05_{-    0.11-    0.16}^{+    0.07+    0.18}$ & $    0.038_{-    0.058-    0.095}^{+    0.042+    0.12}$ & $   -0.0027_{-    0.045-    0.089}^{+    0.043+    0.090}$ &  $    0.041_{-    0.046-    0.079}^{+    0.040+    0.086}$ \\

$\Omega_{m0}$ & $    0.33_{-    0.16-    0.22}^{+    0.09+    0.25}$ & $    0.291_{-    0.036-    0.072}^{+    0.041+    0.068}$ & $    0.293_{-    0.019-    0.047}^{+    0.023+    0.039}$ & $    0.312_{-    0.019-    0.039}^{+    0.020+    0.039}$ & $    0.289_{-    0.017-    0.033}^{+    0.017+    0.033}$  \\

$\sigma_8$ & $    0.84_{-    0.16-    0.21}^{+    0.12+    0.23}$ & $    0.854_{-    0.062-    0.096}^{+    0.044+    0.11}$ & $    0.851_{-    0.037-    0.064}^{+    0.028+    0.074}$ & $    0.818_{-    0.028-    0.049}^{+    0.026+    0.053}$ & $    0.836_{-    0.027-    0.047}^{+    0.024+    0.050}$ \\

$H_0$ & $   66.9_{-    4.3-    9.2}^{+    5.3+    8.7}$ & $   68.3_{-    1.5-    2.5}^{+    1.3+    2.7}$ & $   68.19_{-    0.84-    1.5}^{+    0.71+    1.6}$ & $   67.46_{-    0.76-    1.4}^{+    0.73+    1.5}$ & $   68.47_{-    0.66-    1.3}^{+    0.66+    1.3}$  \\

\hline\hline                             
\end{tabular}
\caption{Observational constraints on the second interaction model IVS2 of (\ref{ivs2}) have been shown using the standard cosmological probes CMB, BAO, Pantheon, RSD, R16, CC and WL. The parameter $\Omega_{m0}$ is the present value of the matter density parameter $\Omega_m = \Omega_c+\Omega_b$, and $H_0$ is in the units of km/sec/Mpc. Here, CB = CMB+BAO, CBP = CMB+BAO+Pantheon, CBPR = CMB+BAO+Pantheon+RSD, CBPRR16CW = CMB+BAO+Pantheon+RSD+R16+CC+WL. }\label{tab:ivs2}                                                                                                   
\end{table*}                                                                                                                     
\end{center}                                                                                                                    
\endgroup 
\begingroup                                                                                                                     
\squeezetable                                   
\begin{center}                              
\begin{table*} 
\begin{tabular}{ccccccccccccccccc}              
\hline\hline                            
Parameters & CG & CBG & CBPG & CBPRG& CBPRR16CWG\\ \hline

$\Omega_c h^2$ & $0.1245_{-    0.0061-    0.012}^{+    0.0067+    0.012}$ & $ 0.1209_{-    0.0065-    0.012}^{+    0.0069+    0.012}$ & $    0.1116_{-    0.0050-    0.012}^{+    0.0064+    0.010}$ & $    0.1173_{-    0.0054-    0.011}^{+    0.0053+    0.010}$ & $    0.1109_{-    0.0047-    0.011}^{+    0.0055+    0.010}$  \\

$\Omega_b h^2$ & $0.02216_{-    0.00016-    0.00029}^{+    0.00015+    0.00031}$ & $0.02235_{-    0.00016-    0.00029}^{+    0.00016+    0.00031}$ & $    0.02225_{-    0.00014-    0.00027}^{+    0.00014+    0.00026}$ & $    0.02224_{-    0.00014-    0.00027}^{+    0.00014+    0.00027}$ & $    0.02227_{-    0.00013-    0.00026}^{+    0.00014+    0.00026}$ \\

$100\theta_{MC}$ & $1.04012_{-    0.00038-    0.00077}^{+    0.00040+    0.00075}$ & $ 1.04055_{-    0.00037-    0.00072}^{+    0.00036+    0.00074}$ & $    1.04095_{-    0.00037-    0.00073}^{+    0.00038+    0.00076}$ & $    1.04060_{-    0.00035-    0.00069}^{+    0.00036+    0.00073}$ & $    1.04101_{-    0.00037-    0.00069}^{+    0.00034+    0.00070}$ \\

$\tau$ & $0.078_{-    0.017-    0.033}^{+    0.017+    0.033}$ & $ 0.089_{-    0.018-    0.034}^{+    0.019+    0.032}$ & $    0.085_{-    0.017-    0.034}^{+    0.019+    0.032}$ & $    0.073_{-    0.017-    0.034}^{+    0.018+    0.034}$ & $    0.066_{-    0.016-    0.034}^{+    0.016+    0.033}$ \\

$n_s$ & $ 0.9723_{-    0.0044-    0.0084}^{+    0.0044+    0.0088}$ & $  0.9780_{-    0.0046-    0.0091}^{+    0.0045+    0.0090}$ & $    0.9754_{-    0.0040-    0.0083}^{+    0.0040+    0.0080}$ & $    0.9748_{-    0.0040-    0.0081}^{+    0.0041+    0.0081}$ & $    0.9753_{-    0.0039-    0.0079}^{+    0.0039+    0.0078}$ \\

${\rm{ln}}(10^{10} A_s)$ & $ 3.100_{-    0.033-    0.064}^{+    0.034+    0.063}$ & $  3.119_{-    0.036-    0.067}^{+    0.035+    0.063} $ & $    3.111_{-    0.033-    0.066}^{+    0.036+    0.063}$ & $    3.086_{-    0.033-    0.066}^{+    0.034+    0.064}$ & $    3.071_{-    0.032-    0.066}^{+    0.032+    0.064}$ \\

$\xi$ & $ -0.028_{-    0.052-    0.091}^{+    0.045+    0.095}$ & $ -0.018_{-    0.052-    0.096}^{+    0.048+    0.096}$ & $    0.048_{-    0.051-    0.086}^{+    0.037+    0.093}$ & $    0.010_{-    0.043-    0.076}^{+    0.041+    0.081}$ & $    0.050_{-    0.044-    0.079}^{+    0.035+    0.083}$  \\

$\Omega_{m0}$ & $    0.336_{-    0.020-    0.038}^{+    0.020+    0.040}$ & $ 0.316_{-    0.020-    0.036}^{+    0.020+    0.038}$ & $    0.288_{-    0.015-    0.034}^{+    0.017+    0.030}$ & $    0.307_{-    0.016-    0.031}^{+    0.016+    0.032}$ & $    0.285_{-    0.014-    0.030}^{+    0.014+    0.027}$ \\

$\sigma_8$ & $   0.812_{-    0.029-    0.055}^{+    0.029+    0.060}$ & $ 0.819_{-    0.032-    0.059}^{+    0.032+    0.056}$ & $    0.857_{-    0.031-    0.053}^{+    0.027+    0.056}$ & $    0.824_{-    0.026-    0.046}^{+    0.024+    0.048}$ & $    0.840_{-    0.026-    0.044}^{+    0.022+    0.048}$ \\

$H_0$ & $   66.24_{-    0.59-    1.1}^{+    0.58+    1.2}$ & $   67.46_{-    0.65-    1.1}^{+    0.60+    1.1}$ & $   68.36_{-    0.55-    1.0}^{+    0.50+    1.1}$ & $   67.61_{-    0.52-    1.0}^{+    0.53+    1.0}$ & $   68.58_{-    0.53-    0.96}^{+    0.47+    0.98}$ \\

\hline\hline                                                                                                                    
\end{tabular}
\caption{Observational constraints on the first interaction model IVS2 of (\ref{ivs2}) have been shown after the inclusion of the gravitational waves data with the standard cosmological probes. The parameter $\Omega_{m0}$ is the present value of the matter density parameter $\Omega_m = \Omega_c+\Omega_b$, and $H_0$ is in the units of km/sec/Mpc. Here, CG = CMB+GW, CBG = CMB+BAO+GW, CBPG = CMB+BAO+Pantheon+GW, CBPRG = CBPR = CMB+BAO+Pantheon+RSD+GW, CBPRR16CWG = CMB+BAO+Pantheon+RSD+R16+CC+WL+GW.} 
\label{tab:ivs2-gw}                                                                                                   
\end{table*}                    
\end{center}                                                                                                                    
\endgroup                                                                                                                       
\begin{figure*}
\includegraphics[width=0.8\textwidth]{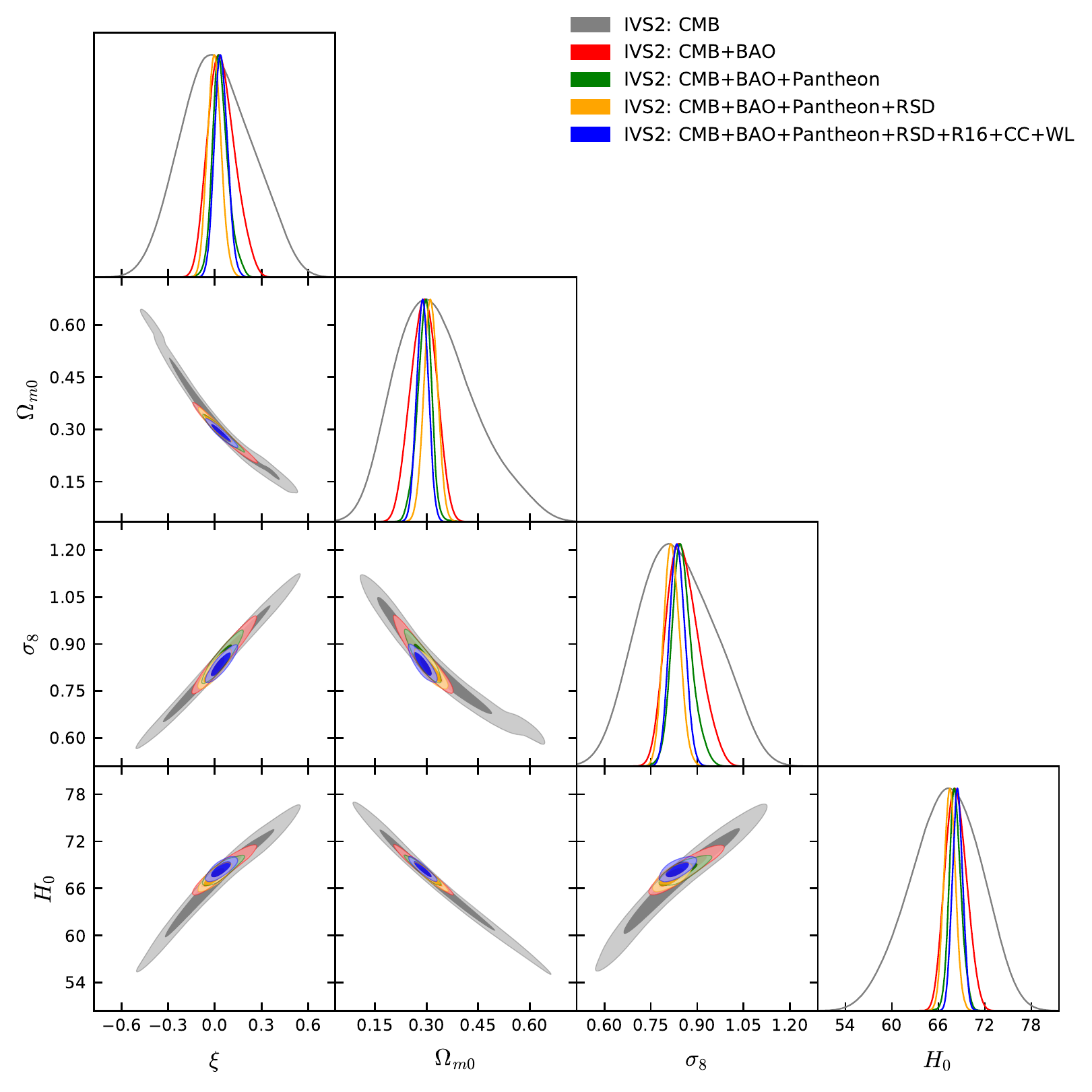}
\caption{The figure for IVS2 shows two different things as follows. It contains the 68\% and 95\% confidence level contour plots between several combinations of the model parameters using only the standard cosmological probes namely CMB, BAO, Pantheon, RSD, R16, CC and WL. Also, it shows the one dimensional marginalized posterior distributions of some selected parameters. We note that here $\Omega_{m0}$ is the present value of $\Omega_m = \Omega_c +\Omega_b$, and $H_0$ is in the units of Km/sec/Mpc. }
\label{fig:real_IVS2}
\end{figure*}
\begin{figure*}
\includegraphics[width=1.0\textwidth]{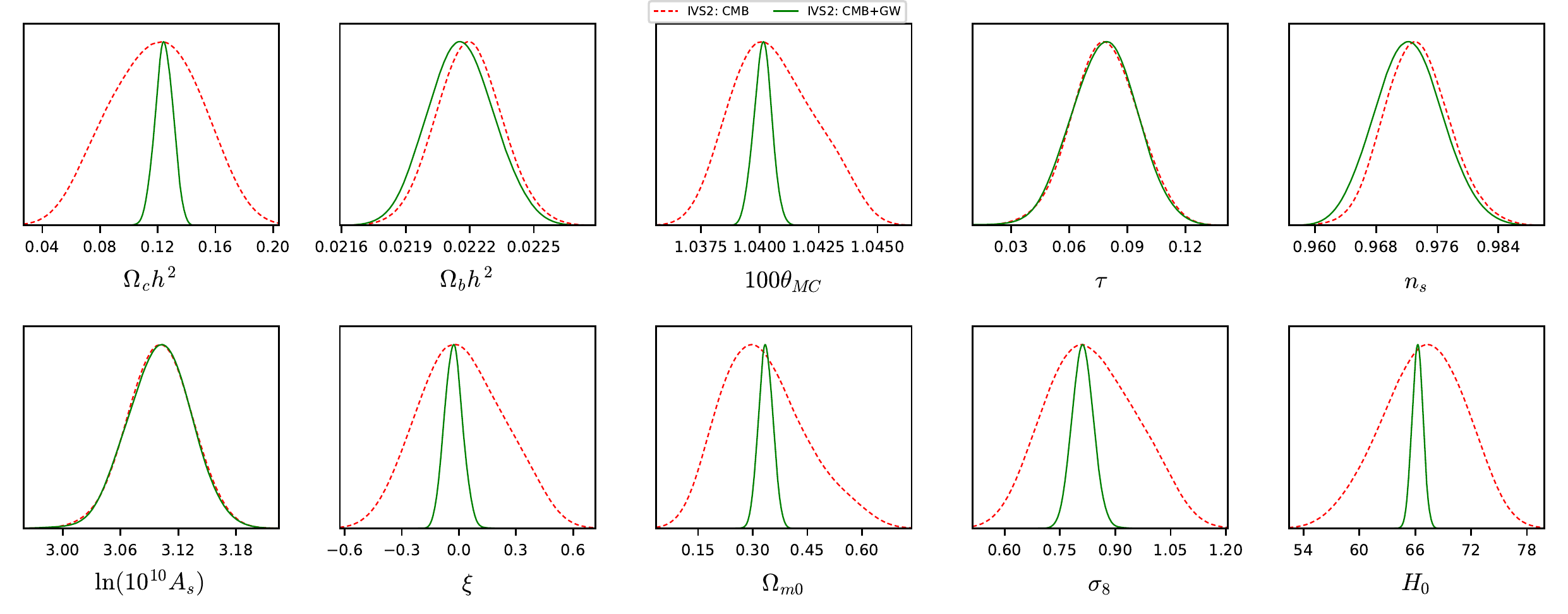}
\caption{We compare the one dimensional marginalized posterior distributions for some selected parameters of the interacting scenario IVS2 considering the observational datasets CMB and CMB+GW. }
\label{fig:1DIVS2-CG}
\end{figure*}
\begin{figure*}
\includegraphics[width=0.42\textwidth]{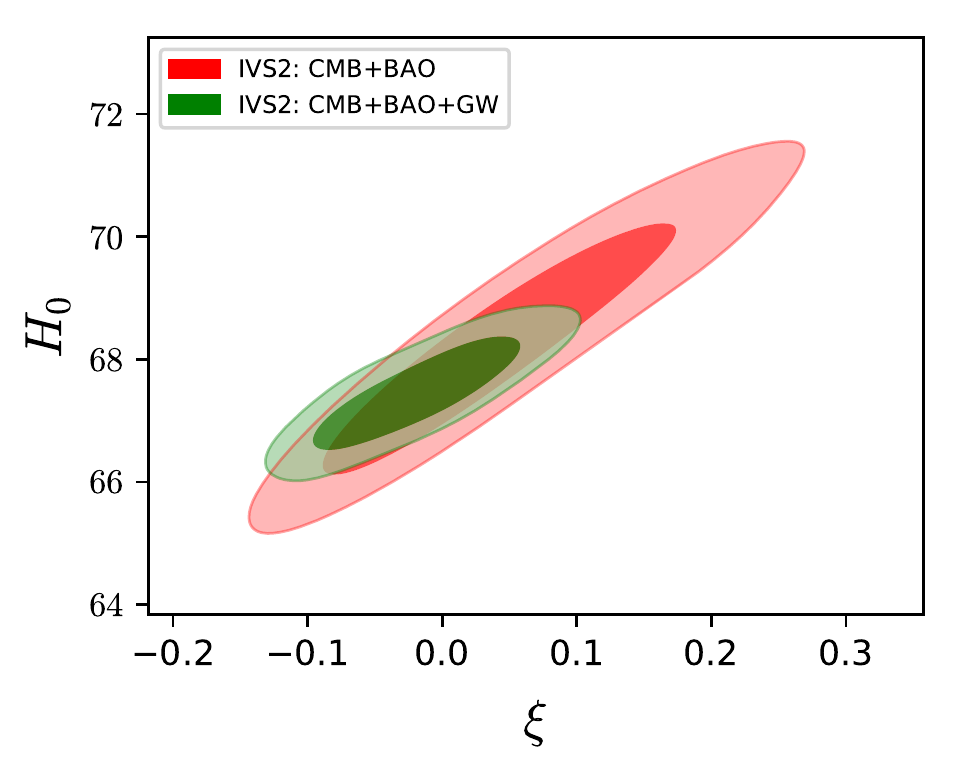}
\includegraphics[width=0.44\textwidth]{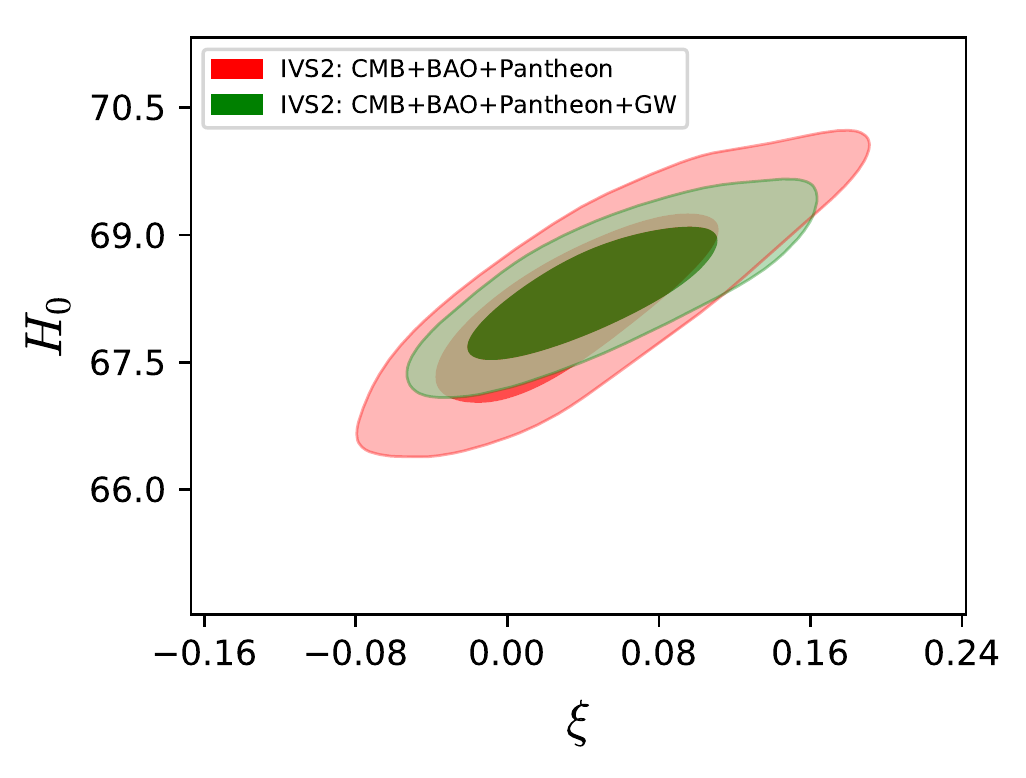}\\
\includegraphics[width=0.42\textwidth]{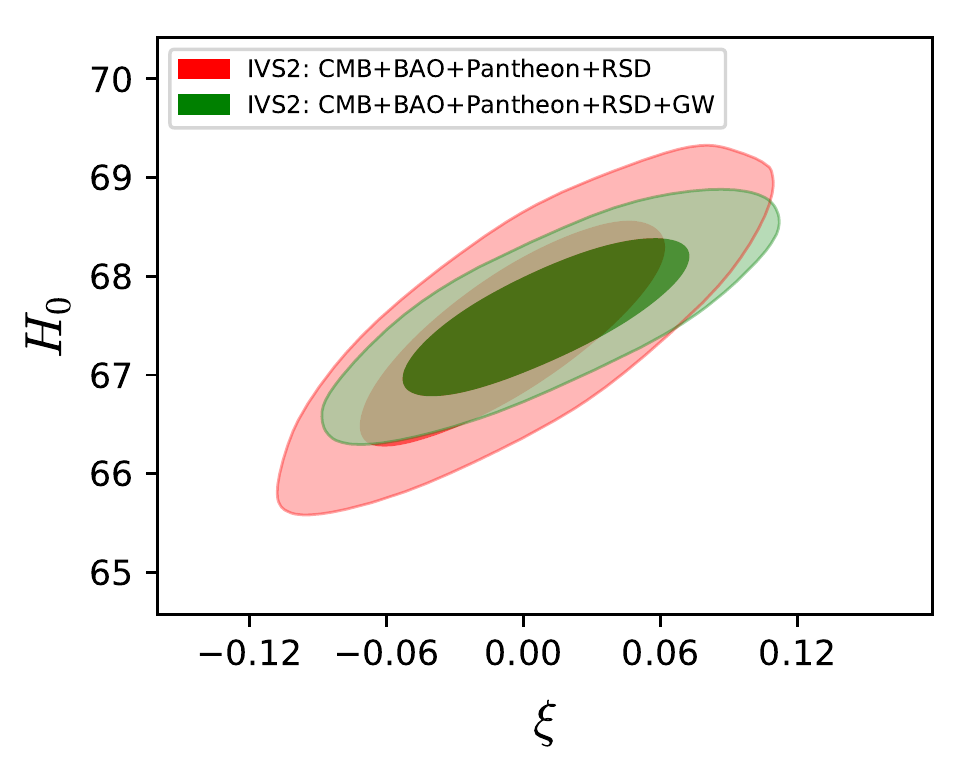}
\includegraphics[width=0.42\textwidth]{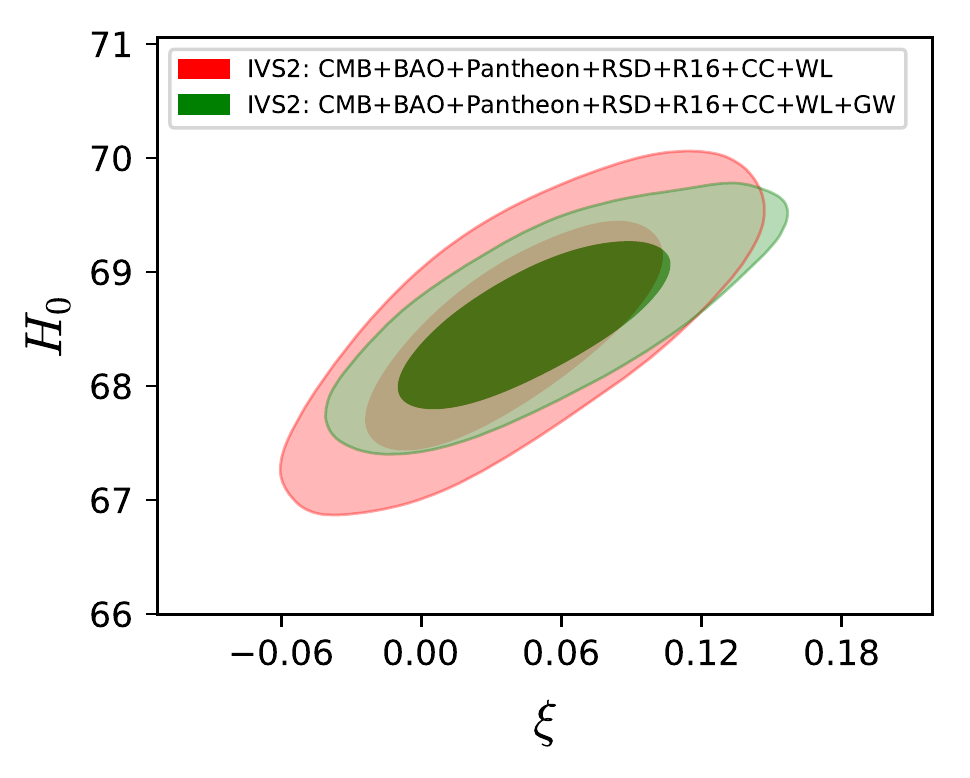}
\caption{In this figure we compare various contour plots of the cosmological parameters for the IVS2 scenario before and after the inclusion of the GWs data. }
\label{fig:2DIVS2}
\end{figure*}

\subsection{IVS1: $Q = 3 H \xi \rho_{x} (t)$}
\label{sec-ivs1}

The observational constraints for this interaction function (equivalently the interaction scenario) using the current cosmological probes, namely CMB, BAO, Pantheon, RSD, R16, CC and WL, have been shown in Table \ref{tab:ivs1-noGW} where specifically we have presented the 68\% and 95\% CL constraints on each free and derived parameters, and in Fig.~\ref{fig:real_IVS1} we show the corresponding triangular plot.  On the other hand, in Table \ref{tab:ivs1-withGW} we have shown the forecasted constraints on the same cosmological parameters of this interaction scenario but after the inclusion of the gravitational waves data that we have simulated using the Einstein Telescope. Thus, the tables, namely Table \ref{tab:ivs1-noGW} and Table \ref{tab:ivs1-withGW}, present an explicit comparisons between the cosmological parameters of this interaction scenario showing how the simulated GW data could affect the cosmological parameters.

From Table \ref{tab:ivs1-noGW} and Fig.~\ref{fig:real_IVS1}, one can see that the cosmological parameters obtained with the introduction of the interactive model are perfectly in agreement within one standard deviation with the bounds obtained from the Planck collaboration assuming the $\Lambda$CDM model, but with weakened constraints, in particular for $H_0$ and $\theta_{MC}$. The reason of this agreement is that the interaction parameter $\xi$ is consistent with a null interaction value. For the CMB data alone, thanks to the larger error bars, one can see that at about $2.3\sigma$ the Hubble constant estimation becomes in agreement to its local measurement by Riess et al. 2016 \cite{Riess:2016jrr} and at about $2.5\sigma$ with Riess et al. 2018 \cite{Riess:2018byc}. The inclusion of BAO to CMB shifts $\xi$, and so $H_0$ that is positively correlated with it (see Fig.~\ref{fig:real_IVS1}), towards higher values with respect to the CMB alone case, and now $H_0$ is in agreement with its local estimations at about $2.2-2.4 \sigma$, even if the error bars are reduced. The inclusion of Pantheon to the former dataset (CMB+BAO) shrinks the constraints, confirming approximately the mean values of $\xi$ and $H_0$, and resulting in an increase of the $H_0$ tension with \cite{Riess:2016jrr} at about $2.7\sigma$ and with \cite{Riess:2018byc} at about $3\sigma$. The addition of RSD to CMB+BAO+Pantheon, shifts $\xi$ back to the zero value, and consequently $H_0$ towards the $\Lambda$CDM value, due to their degeneracy, and the tension at more than $3\sigma$. This happens because of the tension between the RSD dataset and the CMB.    
Finally, the inclusion of the full combination 
i.e., CMB+BAO+Pantheon+RSD+R16+CC+WL (CBPRR16CW), doesn't add any more constraining power to the previous case (CMB+BAO+Pantheon+RSD), and sometimes the error bars are larger, as for example $\sigma_8$, for the tension of CBPR with the R16 dataset.

The inclusion of the simulated GWs to the current standard cosmological probes has a large impact for this interactive model. Looking at  Fig.~\ref{fig:1DIVS1-CG}, we can see that the most affected independent parameters of our analysis by the inclusion of the GWs, are $\Omega_ch^2$, $\theta_{MC}$ and $\xi$, with an effect on the derived parameters $\Omega_m$, $\sigma_8$ and $H_0$. In fact, even if GWs are geometrical probes, and  they are not sensitive to the clustering parameter, however, by breaking the degeneracies between the free and derived parameters in a specific cosmological model, GWs could also improve the constraints on $\sigma_8$, see for instance \cite{Congedo:2018wfn}. That means GWs could indirectly improve the parameters quantifying the large scale behaviour of our universe.  From Table \ref{tab:ivs1-withGW}, one can see that the error bars on the independent cosmological parameters for the CMB  case are expected to have an improvement of a factor more than $2$ by the addition of the GWs data, while the derived parameter $H_0$ significantly decreases of a factor about $4$ compared to its constraints from the only corresponding CMB case (see Table \ref{tab:ivs1-noGW}) reaching an accuracy of 1\% on its determination. We found that CMB+GW performs better with respect to CMB+BAO in Table~\ref{tab:ivs1-noGW}, constraining the cosmological parameters  in a stronger way for this interactive model. This improvement is forecasted to be significant also with respect to the case CMB+BAO, as we can see in Fig. \ref{fig:triIVS1-CBG}, even if slightly less pronounced. In fact, CMB+GW and CMB+BAO+GW are expected to have the same error bars (see Table \ref{tab:ivs1-withGW}), and in other words, BAO will not add any more information in this case. A further improvement is expected with the addition of Pantheon, with a gain of about a factor of $1.2-1.5$ on the previous cosmological parameters by the inclusion of the GWs data to the CMB+BAO+Pantheon combination. And again there is no further improvements with the RSD dataset as we can see by comparing the constraints obtained from the datasets CMB+BAO+Pantheon+GW and CMB+BAO+Pantheon+RSD+GW, summarized in Table \ref{tab:ivs1-withGW}. Finally, we can see an additional improvement  in the constraints of free and derived parameters when one considers the full combination in the last column of Table \ref{tab:ivs1-withGW}.  This improvement results up to a factor of $2$ (one can compare this by looking at the last columns of Table \ref{tab:ivs1-noGW} and Table \ref{tab:ivs1-withGW}. In fact, from the graphical presentation, see Fig. \ref{fig:tri_IVS1-full}, it is clearly manifested. That means the effects of GWs on the concerned cosmological parameters are quite evident.  In particular, we see that it is possible to reach an error of $0.020$ on $\xi$, allowing in principle to detect or have stronger constraints on the interaction parameter, $\xi$, and an accuracy of $0.5\%$ on the Hubble constant $H_0$ shedding light on the current tension between the CMB and the local measurements.

\subsection{IVS2: $Q = 3 H \xi \rho_c (t) \rho_{x} (t) (\rho_c (t) +\rho_{x} (t))^{-1}$}
\label{sec-ivs2}

In this section we describe the results for the IVS2 interactive scenario using different observational such as CMB, BAO, Pantheon, RSD, R16, CC and WL as well as the simulated GWs data from the Einstein Telescope. In Table \ref{tab:ivs2} and Table \ref{tab:ivs2-gw} we have clearly presented  the constraints on all the free and derived parameters (at 68\% and 95\% CL) using the current standard cosmological data and the simulated GWs data, respectively. Thus, Table  \ref{tab:ivs2} and Table \ref{tab:ivs2-gw} effectively summarize this interactive model. 
Also, in Fig. \ref{fig:real_IVS2} we show a  triangular plot for this interactive scenario using only the standard cosmological probes.

From Table~\ref{tab:ivs2} and Fig.~\ref{fig:real_IVS2}, we can see that the cosmological parameters obtained with the introduction of the interactive model are in agreement with the bounds obtained from the Planck collaboration assuming the $\Lambda$CDM model, because $\xi$ is consistent with zero. For the CMB data alone, we find very larger error bars on $\xi$ with respect to the IVS1 model, that corresponds to very large error bars on $H_0$, due to their positive correlation, as we can see from  Fig.~\ref{fig:real_IVS2}. In this case, the Hubble constant is in agreement at about $1.1-1.2\sigma$ with its local measurement by \cite{Riess:2016jrr} and \cite{Riess:2018byc}. The inclusion of BAO to CMB shrinks a lot the constraints shifting $H_0$ slightly towards higher values with respect to the CMB  case alone, in agreement with the same combination of data for the IVS1 model. The inclusion of Pantheon to CMB+BAO halves the error bars on most of the cosmological parameters, restoring the $H_0$ tension with \cite{Riess:2016jrr} at about $2.6\sigma$ and with \cite{Riess:2018byc} at about $3\sigma$. The addition of RSD to CMB+BAO+Pantheon doesn't improve in a significant way the bounds on the cosmological parameters, but shifts the Hubble constant mean value increasing the disagreement at more than $3$ standard deviations. Finally the further inclusion of the full combination we are considering in this work, i.e. R16+CC+WL, brings again the mean value of $H_0$ up, with an accuracy of $1\%$, and the $\tau$ value down, in agreement with Planck 2018 \cite{Aghanim:2018eyx} within one standard deviation.

In Table \ref{tab:ivs2-gw} one can clearly see that the addition of the simulated GWs data to the current cosmological probes significantly improves the parameters space of the model parameters by reducing their error bars.
One can visualize the effects of GWs on the CMB data alone looking at Fig.~\ref{fig:1DIVS2-CG}, or comparing Table \ref{tab:ivs2} and Table \ref{tab:ivs2-gw}.
Also in this case are the independent parameters $\Omega_ch^2$, $\theta_{MC}$ and $\xi$ to be the most affected by the inclusion of the simulated GWs data, improving consequently the derived parameter in Fig.~\ref{fig:1DIVS2-CG}. For example, one can clearly see that the coupling parameter is estimated as $\xi  = 0.02_{-    0.27}^{+    0.22}$ (68\% CL, for CMB alone) and $\xi = -0.028_{-    0.052}^{+    0.045} $ (68\% CL, CMB+GW), which clearly demonstrates the improvement of a factor $5$ in the parameters space for the coupling parameter. Similar effects can be seen from other parameters as well after the inclusion of GWs data to the standard cosmological probes.  CMB+GW is better than CMB+BAO.
We can estimate the effect of the simulated GWs data in the plane $\xi$ vs $H_0$ looking at Fig.~\ref{fig:2DIVS2}, where it is evident for a significant improvement in the parameter space also compared to the full combination of datasets currently available. The inclusion of the simulated GWs data is forecasted to constrain the $\xi$ parameter with an error bar of about $0.040$ and $H_0$ of about $0.5$ respectively.

\section{Concluding remarks}
\label{sec-summary}

Observational data from different astronomical sources have been playing an important role in the understanding of the universe's evolution. Due to successive developments in the observational data our understanding on different cosmological models that we had in the beginning of twenty-first century, has changed a lot at present time. And thanks to the astronomical datasets a large number of cosmological models have been excluded as a consequence and several others have been strictly constrained. Therefore, without any further doubt theory and  observations are complimentary to each other. The main theme of the present work is to examine the constraining power of the cosmological models using a very recent astronomical data, namely the gravitational waves standard sirens.

The GWs have just been detected by LIGO and VIRGO collaborations \cite{ligo01,Abbott:2016nmj,Gw03,Gw04,Gw05, Gw06, Gw07} and probably it is one of the thrilling achievements of modern cosmology possibly after the detection of the accelerating universe. Undoubtedly, with the detection of GWs, a cluster of possibilities for modern cosmological research are now available to us \cite{Creminelli:2017sry,Ezquiaga:2017ekz,Sakstein:2017xjx,
Chakraborty:2017qve,Visinelli:2017bny,Oost:2018tcv,Casalino:2018tcd,Nunes:2018evm,Nunes:2018zot,Mifsud:2019hjw,Qi:2019wwb, Palmese:2019ehe, DiValentino:2017clw,DiValentino:2018jbh}. The understanding of the dark sector's dynamics is one of the possibilities among them.  Being inspired with the detection of GWs, here in this work, we have focused on a specific but general class of cosmological models, namely the interacting DE models  in order to see how future GWs data could constrain these models. 

Thus, to start with we have considered a simple interaction scenario where the pressureless DM fluid interacts with vacuum energy. To illustrate such scenario  we have assumed   
two interaction models shown in equations (\ref{ivs1}) and (\ref{ivs2}) and constrained them with the use of (i) current standard cosmological data and (ii) then with the use of simulated GWs data in presence of the usual cosmological data. The results are summarized in Table \ref{tab:ivs1-noGW} (IVS1; no GWs), Table \ref{tab:ivs1-withGW} (IVS1; with GWs), Table \ref{tab:ivs2} (IVS2; no GWs) and Table \ref{tab:ivs2-gw} (IVS2; with GWs). 

Our analysis is very simple. For both the interaction models we test the corresponding cosmic scenarios with the current standard cosmological data, namely CMB, BAO, Pantheon, RSD, R16, CC, WL, and found a mild interaction in the dark sector which seems to be consistent with the $\Lambda$CDM cosmology. We then add simulated GWs data to the standard cosmological probes and measures the improvements of the cosmological parameters. We find that the addition of GWs data with the standard cosmological probes significantly improves most of the model parameters by reducing their error bars. This is a potential side of the GWs data as its inclusion to standard cosmological probes offers more stringent constraints on most of the cosmological parameters, as also observed in this context, i.e. for IVS1 and IVS2. It deserves to be  mentioned that apart from IVS (i.e. IVS1 and IVS2) studied in this work, the effects of GWs data are equally visible in the interacting dark energy (IDE) scenarios where dark energy is not vacuum, rather, dark energy has a constant equation of state (EoS), $w$, other than $-1$, see for instance a recent study \cite{Yang:2019vni}. Precisely, in Ref. \cite{Yang:2019vni}, the authors find that the constraining power of GWs  is equally visible in some key cosmological parameters, for instance from the measurements of the Hubble constant $H_0$ and the density parameter for matter, $\Omega_{m0}$ (see Table II and Table III of \cite{Yang:2019vni}) one can clearly notice that the inclusion of GWs to CMB significantly reduces the error bars on such parameters, thus, leading to more precise constraints on them. It is perhaps interesting to note that although IDE scenarios have extra degrees of freedom compared to IVS, but, the effects of GWs remain similar. Possibly one might argue that the effects of GWs might be model dependent, here the interaction function, $Q$, however, this is surely a subject of further investigations.

Finally, we comment that one can introduce the massive neutrinos into the interaction scenarios with an aim to check the bounds on the total neutrino mass in presence of the simulated GWs data. We hope to address this issue in a forthcoming work.

\section{Acknowledgments}
The authors sincerely thank the referee whose comments helped us to improve the quality of discussion of the manuscript.  WY was supported by the National
Natural Science Foundation of China under Grants No.  11705079 and No.  11647153. SP acknowledges partial support from the Faculty Research and Development Fund (FRPDF) Scheme of Presidency University, Kolkata, India.  EDV acknowledges the support from the European Research Council in the form of a Consolidator Grant with number 681431. The work of AW was supported in part by the National Natural Science Foundation of China (NNSFC), Grant Nos. 11375153 and 11675145. The authors thank Minghui Du for some important discussions on GW simulations.


\end{document}